\documentclass[11pt,twoside,final]{article}

\usepackage{hyperref}
\hypersetup{colorlinks,linkcolor={blue},citecolor={blue},urlcolor={red}} 

\usepackage{fullpage}

\usepackage[mathscr]{eucal}
\usepackage{mathrsfs}
\usepackage{amsbsy}

\DeclareMathAlphabet{\mathpzc}{OT1}{pzc}{m}{it}

\usepackage{epsf}
\usepackage{fancyheadings}
\usepackage{graphics}
\usepackage{graphicx}
\usepackage{enumitem}
\usepackage{float}

\usepackage{color}

\usepackage{amsthm} \usepackage{amsfonts} \usepackage{amsmath}
\usepackage{amssymb}

\usepackage{bbm}

\usepackage{caption} \usepackage{subcaption}

 \usepackage{macros}
\usepackage[noabbrev,nameinlink]{cleveref}

\newcommand{\figdir}{figures/}

\usepackage{bm}
\usepackage{mathtools}
\usepackage{dsfont}
\usepackage{tikz}
\usetikzlibrary{matrix,decorations.pathreplacing,positioning}
\usepackage{adjustbox}
\usepackage{comment}
\usepackage{placeins}

\setitemize{leftmargin=0.15in,itemsep=0.02in,topsep=0.1in}
\setenumerate{leftmargin=0.2in,itemsep=0.02in,topsep=0.1in}

\newcommand{\noisevar}{\ensuremath{\sigma}}
\newcommand{\sing}{\ensuremath{\gamma}}
\newcommand{\Mstar}{\mathbf{M}^{\star}}
\newcommand{\bM}{\mathbf{M}}
\newcommand{\lvec}{\mathbf{c}_{1}}
\newcommand{\rvec}{\mathbf{c}_{2}}
\newcommand{\bE}{\mathbf{E}}
\newcommand{\bU}{\mathbf{U}}
\newcommand{\bH}{\mathbf{H}}

\newcommand{\bI}{\mathbf{I}}
\newcommand{\bZ}{\mathbf{Z}}
\newcommand{\bSigma}{\mathbf{\Sigma}}
\newcommand{\bX}{\mathbf{X}}
\newcommand{\bY}{\mathbf{Y}}
\newcommand{\bV}{\mathbf{V}}

\newcommand{\estvar}{\ensuremath{\gamma}}
\newcommand{\hatbU}{\widehat{\mathbf{U}}}
\newcommand{\hatbV}{\widehat{\mathbf{V}}}
\newcommand{\EE}{\mathbb{E}}
\newcommand{\clow}{\ensuremath{c_\ell}}
\newcommand{\cupper}{\ensuremath{c_u}}

\newcommand{\SNR}{\ensuremath{\rho_{N,T}}}
\newcommand{\ratio}{\ensuremath{\kappa_{\noisevar}}}

\newcommand{\smallprob}{\ensuremath{O((N+T)^{-10})}}
\newcommand{\spicy}{\ensuremath{\xi_{N,T}}}

\newcommand{\PP}{\mathbb{P}}

\definecolor{yly}{RGB}{125,0,0}

\usepackage[linesnumbered,ruled,vlined]{algorithm2e}

\SetCommentSty{mycommfont}
\usepackage{algorithmic}

\newcommand{\mydefn}{\,=}

\newcommand{\fullcomment}[1]{}

\usepackage[giveninits=true,  
  maxnames = 2,  
  backref=true,
  backend=bibtex,
  style=alphabetic,
  sortlocale=de_DE,
  natbib=true,
  doi=false,
  isbn=false,
  url=false]{biblatex}

\newcommand{\Counter}[2]{\ensuremath{Y_{#1, #2}(0)}}
\newcommand{\Exs}{\EE}

\newcommand{\healthdir}{figures/health_expenditures}
\newcommand{\infantdir}{figures/mortality_rates_infant_wind3}
\newcommand{\uninsurdir}{figures/uninsurance_rates}

\newcommand{\real}{\ensuremath{\mathbb{R}}}

\newcommand{\FourBlockEst}{\mbox{\texttt{FourBlockEst}}}
\newcommand{\FourBlockConf}{\mbox{\texttt{FourBlockConf}}}
\newcommand{\StaggeredConf}{\mbox{\texttt{StaggeredConf}}}

\newcommand{\ObsMatrix}{\ensuremath{\mathbf{Y}}}
\newcommand{\ObsLeft}{\ObsMatrix_{\mathsf{left}}}

\newcommand{\ObsUpper}{\ObsMatrix_{\mathsf{upper}}}

\newcommand{\Mhat}{\ensuremath{\widehat{\bM}}}

\newcommand{\SigStar}{\mathbf{\Sigma}^\star}

\newlength{\mywidth}
\newlist{cdesc}{description}{1}
\setlist[cdesc]{font=\mdseries,itemsep=0.5pt}

\newcommand{\bUhat}{\ensuremath{\widehat{\bU}}}
\newcommand{\bVhat}{\ensuremath{\widehat{\bV}}}

\newcommand{\Emat}{\ensuremath{\mathbf{E}}}
\newcommand{\Ehat}{\ensuremath{\widehat{\Emat}}}

\newcommand{\BlockMatrix}[3]{\ensuremath{ {#1}^{#2, #3}}}
\newcommand{\ObsBlock}[2]{\ensuremath{\BlockMatrix{\ObsMatrix}{#1}{#2}}}
\newcommand{\MstarBlock}[2]{\ensuremath{\BlockMatrix{[\Mstar]}{#1}{#2}}}
\newcommand{\MhatBlock}[2]{\ensuremath{\BlockMatrix{\Mhat}{#1}{#2}}}

\newcommand{\polylog}{\operatorname{polylog}}

\newcommand{\sigmax}{\ensuremath{\sigma_{\max}}}

\newcommand{\tikfig}{tikfig}

\newcommand{\Ustar}{\ensuremath{\mathbf{U}^\star}}
\newcommand{\Vstar}{\ensuremath{\mathbf{V}^\star}}

\newcommand{\InputBlock}[2]{\ensuremath{\ObsMatrix[#1, #2]}}

\newcommand{\bUstar}{\ensuremath{\bU^\star}}
\newcommand{\bVstar}{\ensuremath{\bV^\star}}

\newcommand{\singhat}{\widehat{\sing}}
\newcommand{\Normal}{\mathcal{N}}
\newcommand{\SigHat}{\widehat{\mat{\Sigma}}}

\begin{document}


\begin{center}

  {\bf{\LARGE{ Inference under Staggered Adoption: \\  Case Study
        of the Affordable Care Act}}}
\vspace*{.2in}

{\large{
\begin{tabular}{ccc}
Eric Xia$^{\dagger, \star}$, Yuling Yan$^{\circ, \star}$, and Martin J. Wainwright$^{\dagger, \ddagger}$
\end{tabular}
}}

\vspace*{.2in}

\begin{tabular}{c}
  EECS$^\dagger$ and Mathematics$^{\ddagger}$ \\  
  Laboratory for Information and Decision Systems \\
  Statistics and Data Science Center \\
  Massachusetts Institute of Technology, Cambridge, MA
\end{tabular}

\vspace*{.2in}

\begin{tabular}{c}
  Department of Statistics$^\circ$\\
  University of Wisconsin-Madison, Madison, WI
\end{tabular}

\medskip

\today

\vspace*{.2in}

\begin{abstract}
  Panel data consists of a collection of $N$ units that are observed
  over $T$ units of time.  A policy or treatment is subject to
  staggered adoption if different units take on treatment at different
  times and remains treated (or never at all).  Assessing the
  effectiveness of such a policy requires estimating the treatment
  effect, corresponding to the difference between outcomes for treated
  versus untreated units.  We develop inference procedures that build
  upon a computationally efficient matrix estimator for treatment
  effects in panel data.  Our routines return confidence intervals
  (CIs) both for individual treatment effects, as well as for more
  general bilinear functionals of treatment effects, with prescribed
  coverage guarantees. We apply these inferential methods to analyze
  the effectiveness of Medicaid expansion portion of the Affordable
  Care Act.  Based on our analysis, Medicaid expansion has led to
  substantial reductions in uninsurance rates, has reduced infant
  mortality rates, and has had no significant effects on healthcare
  expenditures.  \let\thefootnote\relax\footnote{$^\star$Equal
  contribution.}
\end{abstract}

\end{center}


\section{Introduction}

Many datasets take the form of panel data, in which a collection of
$N$ units (e.g., individuals, cities, states, countries, companies
etc.), are observed over $T$ time periods.  Panel data arises in a
very wide variety of applications, and the associated methodological
literature is rich (e.g., see the book~\cite{Wool10} and references
therein).  It is frequently the case that some ``treatment'' is
applied to a subset of the units.  Here treatment should be
understood, in a generic sense, as some form of intervention or policy
that is applied.  In the simplest case, the treatment is binary in
nature (e.g., whether or not to vaccinate, or whether or not to join
the EU).  The framework of panel data with binary treatments has been
used to study a plethora of problems, including---among many
others---the effects of tax policy on smoking
rates~\cite{abadie2010synthetic}; the economic benefits of EU
membership for a given country~\cite{Kopinski24poland}; the effects of
``right-to-carry'' gun laws~\cite{donohue2019right}; and the effects
of increases in minimum wage~\cite{card1993minimum}.

A fundamental issue underlying analysis of such panel data is the
manner in which the treatment is adopted.  Most straightforward is the
randomized controlled trial, in which a randomly chosen subset of the
units are given treatment at a common time, with all other units
remaining untreated throughout time.  In contrast, the focus of this
paper is a more challenging setting: each unit can choose whether or
not to adopt the treatment, and moreover can choose a time at which to
do so. Furthermore, when a unit adopts treatment, they continue to do
so until the end of the panel time period. This set-up is known as
\emph{staggered adoption}.  It leads to statistical inference problems
that are challenging, both because of the observational nature of the
data, and because of the differing adoption times.  The Affordable
Care Act (ACA) provides an archetypal example of panel data with
staggered adoption.  Here there are $N = 50$ states in total, and we
can measure various features of a given state over a period of $T$
time units.  States can choose whether or not to take on the expanded
Medicaid eligibility provided by the ACA.  From its inception in 2010
through 2024, forty states have chosen to expand Medicaid eligibility
at different times; see~\Cref{fig:expansion} for a graphical
illustration, and~\Cref{sec:ACA} for additional background.  On the
methodological side, there are a wide variety of approaches to
statistical inference with panel data under staggered adoption,
including synthetic controls
(e.g.,~\cite{abadie2003economic,abadie2015comparative,
  li2020statistical, doudchenko2016balancing, ben2021augmented});
linear panel models or fixed effects regression
(e.g.,~\cite{Imai_Kim_2021, athey2006identification}); and approaches
based on matrix completion (e.g.,~\cite{athey2021matrix,
  abadie2024doubly,yan2024entrywise}).  We examine the latter class of
methods in this paper.

More specifically, this paper makes two primary contributions to the
growing literature on panel data and staggered adoption.  The first is
methodological: building upon our recent work~\cite{yan2024entrywise}
on a low-rank matrix-based estimator for treatment effects, we show
how to use its outputs to construct confidence intervals.  We develop
inferential procedures that are sufficiently flexible to handle
heteroskedastic noise, and applicable to both individual treatment
effects (ITEs), as well as to a more general notion of treatment
effect as defined by a bilinear function.  Our second contribution is
to apply these inferential procedures so as to evaluate the
effectiveness of Medicaid expansion component of the the Affordable
Care Act (ACA).  In particular, we study its effect on a variety of
outcomes, including uninsurance rates, infant mortality, and
expenditures.  Our inferential procedures allow for fine-grained
probing of these effects at the individual state level over each time
period. Finally, we have implemented a Python package of our method
for practitioners; see \url{https://github.com/Facta-Non-Verba/CAST_panel}
for a guide on its use, as well as code and data for replicating our results.

The remainder of the paper is organized as
follows. \Cref{sec:background} provides some context on the ACA and
its significance, and then lays out the framework of panel data with
staggered adoption, including the treatment effects to be estimated.
\Cref{sec:methods} reviews our matrix-based estimator of treatment
effects, and provides simple schemes to compute confidence intervals;
we state some informal guarantees on its coverage properties, with all
mathematical details deferred to the Appendices.
In~\Cref{sec:results}, we use these inferential routines to analyze
the causal effects of the Medicaid expansion in the ACA. We summarize
and discuss future avenues of research in~\Cref{sec:conclusion}.


\section{Background and problem formulation}
\label{sec:background}

In this section, we provide some background as well as a more precise
formulation of panel data with staggered adoption.  More specifically,
in~\Cref{sec:ACA} we provide relevant background on the Affordable
Care Act and a brief overview of some related work.
\Cref{sec:causal-setup} is devoted to a more precise description of
the framework of panel data with staggered adoption.

\subsection{The Affordable Care Act}
\label{sec:ACA}

Among all developed nations, residents of the United States are
expected (on average) to live the shortest lives, endure the most
number of chronic health conditions, and experience the highest rates
of infant and maternal mortalities.  All of these negative outcomes
occur despite the US spending the highest proportion of its GDP on
healthcare, and nearly double the average of OECD
nations~\cite{commonwealth2023}. A major contributing factor is lack
of healthcare access due to an inability to pay. For instance, in
2024, roughly 21\% of Americans reported that they skipped seeking
treatment due to costs; moreover, this rate of those skipping care
nearly triples to 61\% among those who do not have
insurance~\cite{kff2024}. In 2022, the US Census reported that nearly
26 million Americans lack health insurance~\cite{kff2023}.

\begin{figure}[h!]
  \begin{center}
  \includegraphics[scale=0.35]{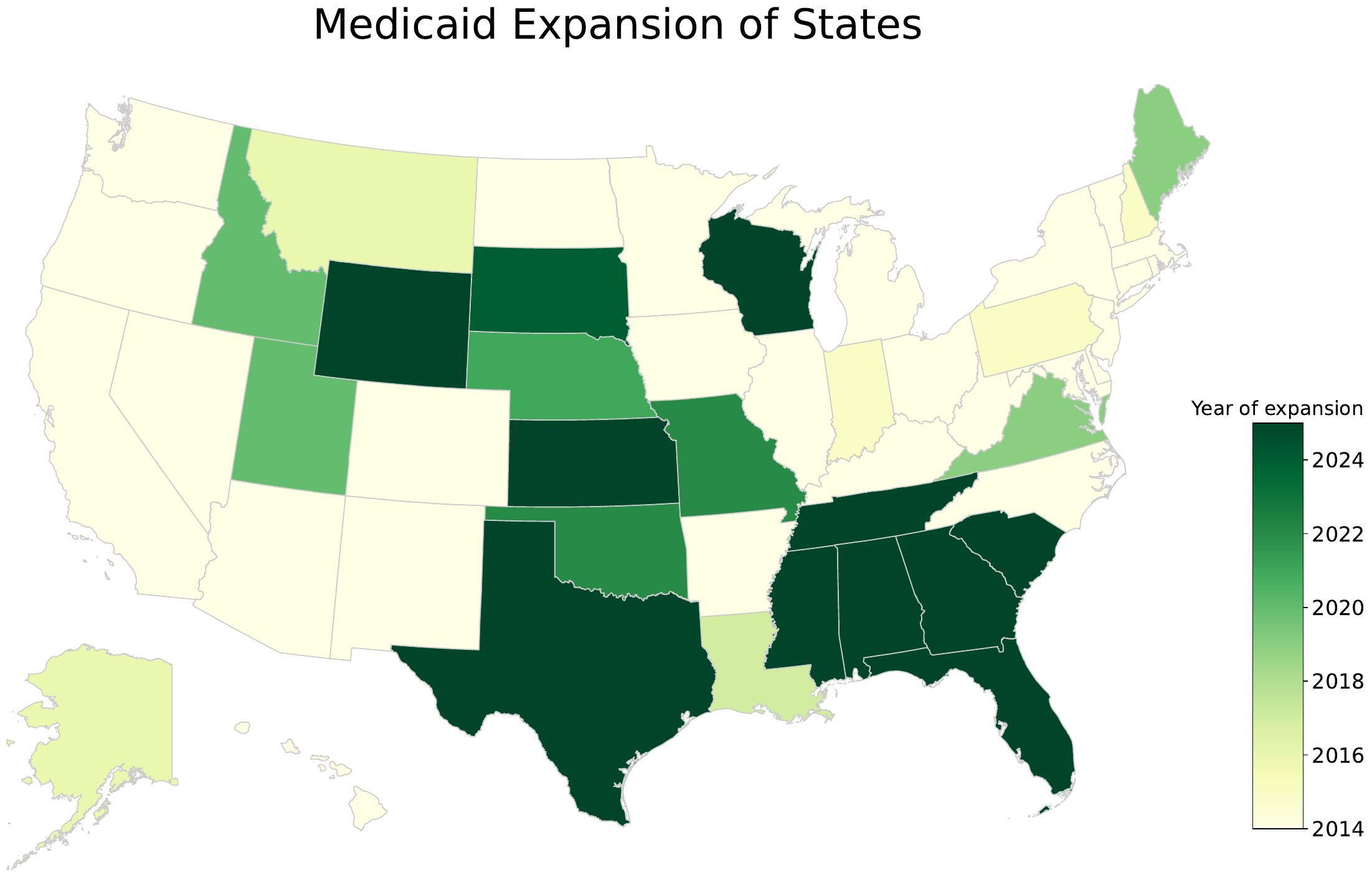}
  \caption{Adoption of Medicaid expansion by state. Darkness of colors
    indicates the time of adoption, with earlier adoption in lighter
    colors.  The 10 states marked in very dark green (i.e., Alabama,
    Florida, Georgia, Kansas, Mississippi, South Carolina, Tennessee,
    Texas, Wisconsin and Wyoming) have not adopted the expansion.}
    \label{fig:expansion}
  \end{center}
\end{figure}

The first major national reform to address these issues was the ACA,
passed in 2010. Its primary goal is to ensure that more Americans have
access to affordable health insurance, without it being tied to one's
employment. One key component of the act was the expansion of
eligibility for Medicaid, a government program established in 1965 to
provide insurance to low income individuals and households. Medicaid
provides coverage individuals that live under the federal poverty
line, and in 2014, it covered nearly 25\% of
Americans~\cite{medicaidballot}. A key component of the ACA was an
incentive to expand Medicaid coverage: the federal government provides
funds to defray the additional costs of states choosing to expand
coverage.  This expanded coverage applies to households whose incomes
amounted up to 138\% of the federal poverty level~\cite{acaballot}. As
of 2024, 40 states have adopted the Medicaid expansion, whereas many
of the remaining states are considering the
expansion~\cite{nyt2014kansas}; see~\Cref{fig:expansion} for a
graphical illustration of the current adoption.  Nonetheless, the ACA
has remained politically contentious.  In 2018, there were significant
efforts made to repeal it, with various arguments made both in
opposition and support.  With the return of a Republican presidential
administration avowing to trim the government in 2025, it is quite
possible that repeal of ACA will again be raised.  For these reasons,
understanding its effects, both on health outcomes and expenditures,
has a timely importance.

Proponents of Medicaid expansion often point to statistics such as
improved access to care and health outcomes in states that chose to
expand access to Medicaid compared to those that did not opt in. For
instance, Miller et.~al.~\cite{miller21medicaid} estimates that
between 2014 and 2017, in the states that chose to expand Medicaid,
there were approximately 19,200 fewer deaths among low-income adults
in the age group 55--64; moreover, they estimate that there were
approximately 15,600 preventable deaths in states that chose not to
opt in.  States that have not taken on expanded eligibility, compared
to those that did opt in, have uninsurance rates that are nearly twice
as large~\cite{kff2024covgap}. However, opponents often argue that the
ACA ``has not stopped the stampede of rising health care costs'', and
that ``nearly 30 millions Americans [are] still
uninsured''~\cite{moore2018obamacare}. There is past work on analyzing
the causal effect of Medicaid expansion: in 2008, Oregon enacted a
lottery that provided Medicaid coverage to previously ineligible
randomly selected low-income adults~\cite{finkelstein2012oregon,
  baicker2013oregon}. These studies documented mixed effects of such
expansion: over the period 2008--2010, Medicaid expansion reduced
uninsurance rates by 25\%, led better self-reported health outcomes
and higher utilization of healthcare, but produced no improvements in
measured physical health.  Our paper builds upon these results, in
particular by developing inferential methods that can provide
treatment effect estimates for observational data, and can be targeted
at the state level, thereby helping identify which states benefit the
most (or the least) from Medicaid expansion.

\subsection{Panel data with staggered adoption}
\label{sec:causal-setup}

We now turn to a more precise formulation of the problem of estimating
treatment effects in panel data under staggered adoption.  In a panel
data model, there are a total of $\numobs$ units observed over
$\numtimes$ time periods.  For each unit index $i  \in  [\numobs] \defn
\{1, \ldots, \numobs \}$ and time $t  \in  [\numtimes] \defn \{1,
\ldots, \numtimes \}$, we observe a scalar outcome $Y_{i,t}$.  There
is an underlying binary treatment, and for each unit $i  \in 
[\numobs]$, we define $t_{i}  \in  [\numtimes] \cup \{+\infty \}$ to be
the time at which unit $i$ adopted the treatment, with $t_i = +\infty$
indicating non-adoption.
\begin{figure}[h]
  \centering
  \input{\tikfig/tikfig_matrix_pattern.tex}
  \caption{An example of panel data under staggered adoption design,
    with units sorted according to the first time of treatment. The
    labels C and A refer to ``control'' and ``adopted'' (i.e.,
    treated) respectively.} \label{fig:intro}
\end{figure}
With this set-up, we can represent the full collections of
observations $\{Y_{i,t}, i \in [N], t \in [T] \}$ as an $N \times T$
matrix, as shown in \Cref{fig:intro}.  Each entry of this matrix can
be labeled with a ``C'' for control (or untreated) or a ``A'' for
adopted (or treated), with the transition between ``C'' and ``A'' in
each row demarcated by the time $t_i$.

\paragraph{Counterfactual outcomes:} 
Let us now introduce the notion of counterfactual outcomes, following
past work on this type of problem (e.g.,~\cite{borusyak2024revisiting,
  eli21synthcontstag,yan2024entrywise}).  We define a random matrix
$\{\Counter{i}{t}, i \in [N], t \in [T] \}$ corresponding to the
potential outcome of unit $i$ and time $t$ if it never undergoes
treatment.  We assume that for each $i \in [N]$ and time unit $t <
t_i$, we observe $Y_{i,t} = \Counter{i}{t}$, or equivalently
\begin{align*}
Y_{i,t} & = \underbrace{\Exs[\Counter{i}{t}]}_{\defn M^\star_{i,t}} +
\noise_{i,t},
\end{align*}
where $\noise_{i,t}$ is (by definition) a zero-mean random variable.
By following this set-up, we are adopting the \emph{no anticipation
assumption}, meaning that observations $Y_{i,t}$ for times $t < t_i$
have the same statistical structure whether or not the treatment is
applied in the future.  On the other hand, our analysis \emph{does
not} impose any assumptions on the statistical structure of any
observation $Y_{i,t}$ for $t \geq t_i$.

For any index $i \in [N]$ and time $t \geq t_i$, we can define an
individual treatment effect, with the full collection corresponding to
unit-time indexed stochastic process
\begin{align}
\label{eqn:ITE}
\TE_{i,t} = Y_{i, t} - \EE[\Counter{i}{t}] \equiv Y_{i,t} - M^\star_{i,t},
\qquad \text{for $i \in [\numobs]$, and $t \geq t_i$.}
\end{align}
In words, the \emph{individual treatment effect} $\TE_{i, t}$, or ITE
for short, is the difference between the observed outcome $Y_{i,t}$
for a treated unit $i$ at time $t$ against the \emph{average
counterfactual outcome} $M^\star_{i,t}$ associated with never adopting the
treatment.  Thus, for the unit $i$ who adopts treatment at time $t_i$,
the ITE $\TE_{i,t}$ for $t \geq t_i$ can be interpreted as the causal
effect of adopting treatment. Also of interest is the 
averaged treatment effect on treated (ATET) at time $t$ as:
\begin{align}
\label{eqn:ATET}
\widebar{\tau}_{t} \defn \frac{1}{\sum_{i} \mathbf{1}(t \geq t_i)} \sum_{i} \tau_{i, t} \cdot \mathbf{1}(t \geq t_i).
\end{align}
Recall that $Y_{i,t}$ is an observed
quantity, so that all challenges associated with inference lie in the unknown counterfactual outcomes $M^\star_{i,t} =
\Exs[\Counter{i}{t}]$.

The main methodological contribution of this paper is to develop
procedures for computing confidence intervals (CIs) for both any
individual treatment effect $\TE_{i,t}$, as well as (more generally)
for weighted bilinear functions defined on the treatment effect
process, of which the ATET is a special case. See~\Cref{AppBilinear}
for details of the bilinear set-up.  These inference routines build
upon a matrix-based estimator of the treatment effect from our past
work~\cite{yan2024entrywise}; this estimator is predicated upon an
additional structural condition on the matrix of mean potential
outcomes, which we now describe.

\paragraph{Low-rank factor model:}
Let $\Mstar \in \real^{N \times T}$ be the full matrix of average
potential outcomes for the control---that is, with entries
$M^\star_{i,t} = \Exs[\Counter{i}{t}]$.  The inferential routines of
this paper build upon an estimation algorithm~\cite{yan2024entrywise}
designed to exploit low-rank structure in the matrix $\Mstar$.  In
particular, we assume that it has a rank $r < \min \{N, T \}$, and so
can be decomposed as
\begin{align}
\label{EqnLowRank}  
  \Mstar & = \bU^\star \bSigma^\star (\bV^\star)^\top \; = \;
  \sum_{j=1}^r \sing_j^\star \bU_j^\star (\bV_j^\star)^\top
\end{align}
Here $\bSigma^\star \defn \mathsf{diag} \{ \sing_1^\star, \ldots,
\sing_r^\star \}$ is a diagonal matrix containing the singular values
in non-increasing order $\sing_1^\star \geq \ldots \geq \sing_r^\star
> 0$.  The matrices $\bU^\star \in \RR^{N \times r}$ and $\bV^\star
\in \RR^{T \times r}$ contain the singular vectors, and can be written
in terms of their columns as
$\bU^\star = \begin{bmatrix} \bU^\star_1 & \cdots \bU^\star_r
\end{bmatrix}$ and
$\bV^\star = \begin{bmatrix} \bV^\star_1
  & \cdots \bV^\star_r
\end{bmatrix}$.
Low-rank assumptions of the type~\eqref{EqnLowRank} originated in the
literature on factor models for panel data~\cite{bai2002determining,
  bai2003inferential, bai2009panel}.  For panel data with treatments,
there are various methods for estimating treatments that rely on a
low-rank assumption; for example, see the
papers~\cite{abadie2010synthetic, eli21synthcontstag} for methods
based on synthetic controls, and the
paper~\cite{borusyak2024revisiting} for a method based on
differences-in-differences approaches. In particular, the literature on estimating ITEs in staggered adoption
settings is quite limited: previous work~\cite{eli21synthcontstag, choi2023matrix}
that study staggered adoption only provide guarantees for estimating
average treatment effects, and other works~\cite{bai2021matrix} that provide ITE guarantees
only apply to the four-block design (c.f.~\Cref{SecReduction}). Our approach in this paper
falls within the class of matrix completion methods
(e.g.,~\cite{athey2021matrix, abadie2024doubly,yan2024entrywise}). 

A final comment to close this section: in general, without some type
of structural condition, the treatment effect~\eqref{eqn:ITE} is
unidentifiable, since it depends on quantities---namely, $M^\star_{i,t} =
\EE[\Counter{i}{t}]$ for $t \geq t_i$---for which we have no
observations.  The low rank assumption~\eqref{EqnLowRank} is one way
in which to enforce identifiability; it provides a strong coupling
between these unobserved quantities and the other matrix entries
$M^\star_{i,t}$ for $t < t_i$, for which our observations are directly
relevant.


\section{Inferential routines for treatment effects}
\label{sec:methods}

We now turn to methods for estimation and inference of treatment
effects within our set-up.  Recall from equation~\eqref{eqn:ITE} the
definition of the individual treatment effect $\TE_{i,t}$.  Since the
quantity $Y_{i,t}$ is observed, estimating and returning confidence
intervals for $\TE_{i,t}$ is equivalent to doing so for the unknown
mean counterfactual outcome $M^\star_{i,t}$.  For this reason, our
discussion focuses on the matrix $\bM^\star$ of these mean outcomes,
but with the understanding that we can move freely back to the
treatment effects.

In addition to confidence intervals (CIs) for the individual treatment
effect~\eqref{eqn:ITE}, the methods to be described here can also be
used to compute CIs for more general objects.  For instance, given an
arbitrary vector $w \in \real^N$, we can do so for the
\textit{weighted treatment effect on treated}, given by
\begin{align}
\label{eqn:WTE}
\tau_{w, t} \defn \sum_{i=1}^\numobs w_i \tau_{i, t} \mathbf{1}(t_i
\geq t).
\end{align}
Even more generally, we can do so for an arbitrary bilinear form of
the matrix $\bM^\star$; see~\Cref{SecCoverage} and~\Cref{AppBilinear}
for further discussion.

In the remainder of this section, we first describe our strategy
reducing a general staggered design to a sequence of simpler
four-block problems (\Cref{SecReduction}).  We describe how to perform
estimation and inference in this simpler setting, before describing
our complete algorithm in~\Cref{SecStaggered}.  Finally,
\Cref{SecCoverage} is devoted to discussion of coverage guarantees for
our confidence intervals, with the technical details deferred to the
appendices.

\subsection{Reduction to four-block design}
\label{SecReduction}

We tackle the general problem by reducing it to a sequence of simpler
problems, ones that exhibit a structure that we refer to as 4-block
design.
\begin{figure}[h]
  \centering 	\begin{tabular}{cc}
		\begin{adjustbox}{valign=t, raise=55mm}
		\begin{tikzpicture}
			\matrix (m) [matrix of math nodes, 
			nodes ={draw,  minimum height =0.75cm,  minimum width =1.1cm,  anchor =center}, 
			column sep =-\pgflinewidth,  row sep =-\pgflinewidth]{
				|[fill =gray!20]| \bM_{1, 1} &  |[fill =gray!20]| \bM_{1, 2} &  |[fill =gray!20]| \bM_{1, 3} & |[fill =gray!20]| \bM_{1, 4} & |[fill =gray!20]| \bM_{1, 5} &  |[fill =gray!20]| \bM_{1, 6} \\
				|[fill =gray!20]| \bM_{2, 1} &  |[fill =gray!20]| \bM_{2, 2} &  |[fill =gray!20]| \bM_{2, 3} & |[fill =gray!20]| \bM_{2, 4} & |[fill =gray!20]| \bM_{2, 5} & \,  \\
				|[fill =gray!20]| \bM_{3, 1} & |[fill =gray!20]| \bM_{3, 2} & |[fill =gray!20]| \bM_{3, 3} &  |[fill =gray!20]| \bM_{3, 4} & \,  & \,  \\
				|[fill =gray!20]| \bM_{4, 1} & |[fill =gray!20]| \bM_{4, 2} &  |[fill =gray!20]| \bM_{4, 3} & \,  & \,  & \,  \\
				|[fill =gray!20]| \bM_{5, 1} & |[fill =gray!20]| \bM_{5, 2} & \,  &  \, & \,  & \,  \\
				|[fill =gray!20]| \bM_{6, 1} & \,  & \,  & \,  & \,  & \,  \\
			};
			
			\foreach \n in {1, ..., 6} {
				\node[above] at (m-1-\n.north) {$T_{\n}$};
			}
			
			\foreach \n in {1, ..., 6} {
				\node[left] at (m-\n-1.west) {$N_{\n}$};
			}
			
	\end{tikzpicture}
	\end{adjustbox}  & 
		\begin{tikzpicture}
			\matrix (m) [matrix of math nodes, 
			nodes={draw,  minimum height=0.75cm,  minimum width=1.1cm,  anchor=center,  font=\small}, 
			column sep=-\pgflinewidth,  row sep=-\pgflinewidth]{
				|[fill=red!20]| \,  & |[fill=red!20]| \,  & |[fill=blue!20]| \,  & |[fill=blue!20]| \,  & |[fill=gray!20]| \,  & |[fill=gray!20]| \,  \\
				|[fill=red!20]| \,  & |[fill=red!20]| \,  & |[fill=blue!20]| \,  & |[fill=blue!20]| \,  & |[fill=gray!20]| \,  & \,  \\
				|[fill=red!20]| \,  & |[fill=red!20]| \,  & |[fill=blue!20]| \,  & |[fill=blue!20]| \,  & \,  & \,  \\
				|[fill=yellow!20]| \,  & |[fill=yellow!20]| \,  &  |[fill=gray!20]| \,  & \,  & \,  & \,  \\
				|[fill=yellow!20]| \,  & |[fill=yellow!20]| \,  & \,  & \bm{?} & \,  & \,  \\
				|[fill=gray!20]| \,  & \,  & \,  & \,  & \,  & \,  \\
			};
			
			\draw[decorate, decoration={brace, amplitude=6pt, raise=1pt}]
			(m-1-1.north west) -- (m-1-2.north east) node [black, midway, yshift=13pt] {$\bar{T}_1$};
			\draw[decorate, decoration={brace, amplitude=6pt, raise=1pt}]
			(m-1-3.north west) -- (m-1-4.north east) node [black, midway, yshift=13pt] {$\bar{T}_2$};
			\draw[decorate, decoration={brace, amplitude=6pt, raise=1pt, mirror}]
			(m-6-1.south west) -- (m-6-4.south east) node [black, midway, yshift=-13pt] {$\bar{T}$};
			
			\draw[decorate, decoration={brace, amplitude=6pt, raise=1pt, mirror}]
			(m-1-1.north west) -- (m-3-1.south west) node [black, midway, xshift=-15pt, yshift=-8pt, anchor=south] {$\bar{N}_1$};
			\draw[decorate, decoration={brace, amplitude=6pt, raise=1pt, mirror}]
			(m-4-1.north west) -- (m-5-1.south west) node [black, midway, xshift=-15pt, yshift=-8pt, anchor=south] {$\bar{N}_2$};
			\draw[decorate, decoration={brace, amplitude=6pt, raise=1pt}]
			(m-1-6.north east) -- (m-5-6.south east) node [black, midway, xshift=13pt, yshift=-8pt, anchor=south] {$\bar{N}$};
			
			\draw[line width=0.5mm] (m-1-1.north west) rectangle (m-5-4.south east);
			\draw[line width=0.5mm] (m-1-1.north west) rectangle (m-3-4.south east);
			\draw[line width=0.5mm] (m-1-1.north west) rectangle (m-3-2.south east);
			\draw[line width=0.5mm] (m-1-1.north west) rectangle (m-5-2.south east);
		\end{tikzpicture}  
	\end{tabular}
	
  \caption{Left: After suitable re-ordering of the rows, panel data
    under staggered adoption can be converted to the ``staircase''
    form shown here.  Right: The staircase pattern can be decomposed
    into a number of smaller ``four-block'' designs, as shown here.
    Our procedure proceeds by first extracting these four-block
    designs, and then performing inference on each.}
  \label{fig:alg-general}
\end{figure}
\Cref{fig:alg-general} provides an illustration of this structure and
its relevance to the general problem.  In the right panel, we show the
generic structure of panel data under staggered adoption; by suitably
re-ordering the rows, we can always convert it to a matrix with a
staircase pattern, as shown here, that separates the treated and
untreated blocks.  A matrix in this staircase pattern can be further
decomposed into a collection of four-block matrices; the left panel
exhibits the extraction of this structure.  See~\Cref{AppPartition}
for a more formal description of this partitioning procedure.

More formally, a panel data problem with four-block structure takes
the following form.  The $N$ units can be divided into two subgroups:
a subset of size $N_1$ that are never exposed to the treatment, with
all the other $N_2 \coloneqq N - N_1$ units receiving the treatment at
the same time $T_1+1$. Consequently, both the potential outcome matrix
$\Mstar$ and the observed matrix $\ObsMatrix$ can be partitioned as
\begin{align}
\label{Eqn4Block}  
\Mstar = \left [ \hspace{-0.5ex} \begin{array}{cc} \Mstar_{a} \in
    \RR^{N_1 \times T_1} & \Mstar_{b} \in \RR^{N_1 \times T_2}
    \\ \Mstar_{c} \in \RR^{N_2 \times T_1} & \Mstar_{d} \in \RR^{N_2
      \times T_2}
  \end{array}  \hspace{-0.5ex} \right ], \quad
\ObsMatrix = \left [ \hspace{-0.5ex} \begin{array}{cc} \ObsMatrix_{a}
    & \ObsMatrix_{b} \\ \ObsMatrix_{c} & \bm{?}
  \end{array}   \hspace{-0.5ex} \right ] =  \left [  \hspace{-0.5ex} \begin{array}{cc} \Mstar_{a} + \bE_a & \Mstar_{b} + \bE_b \\ \Mstar_{c} + \bE_c
    & \bm{?}
  \end{array}   \hspace{-0.5ex} \right ]
\end{align}
where $T_2 = T - T_1$, and $\bE_a$, $\bE_b$, $\bE_c$ are the noise in
the observation at each block. The blocks with a question mark are
unobserved (counterfactual) outcomes.

\subsubsection{A key subroutine:  Four-block matrix estimation}
\label{sec:four-block}

Given a matrix in four-block form, our general procedure exploits a
sub-routine, described formally as Algorithm \FourBlockEst, designed
to estimate the unknown block of the matrix $\Mstar$.  This particular
procedure was proposed in the factor model
literature~\cite{bai2021matrix}, and is also a core sub-routine of our
matrix estimator~\cite{yan2024entrywise} for the more general setting
of staggered adoption.

\begin{algorithm}[h!]
  \DontPrintSemicolon \SetNoFillComment
  \textbf{Input:} Data matrix $\ObsMatrix \in \real^{N\times T}$,
  dimension parameters $N_1$ and $T_1$, rank $r$ \\
  \tcc{Step 1: Subspace Estimation} Compute the truncated rank-$r$ SVD
  $(\bU_{\mathsf{left}}, \bSigma_{\mathsf{left}},
  \bV_{\mathsf{left}})$ of $\ObsLeft$. \\
  Partition $\widehat{\bU}\coloneqq \bU_{\mathsf{left}}$ into two
  submatrices $\widehat{\bU}_{1}$ and $\widehat{\bU}_{2}$, where
  $\widehat{\bU}_{1} \in \mathbb{R}^{N_{1}\times r}$ consists of its
  top $N_{1}$ rows and $\widehat{\bU}_{2} \in \mathbb{R}^{N_{2}\times
    r}$ consists of its bottom $N_{2}$ rows. \\
  \tcc{Step 2: Matrix Denoising} Compute the truncated rank-$r$ SVD $(
  \bU_{\mathsf{upper}}, \bSigma_{\mathsf{upper}},
  \bV_{\mathsf{upper}}) $ of $\ObsUpper$. \\
  Partition $\widehat{\bV}\coloneqq \bV_{\mathsf{upper}}$ into two
  submatrices $\widehat{\bV}_{1}$ and $\widehat{\bV}_{2}$, where
  $\widehat{\bV}_{1} \in \mathbb{R}^{T_{1}\times r}$ consists of its
  top $T_{1}$ rows and $\widehat{\bV}_{2} \in \mathbb{R}^{T_{2}\times
    r}$ consists of its bottom $T_{2}$ rows. \\
  Compute the matrix estimate $\Mhat_{b} \coloneqq
  \bU_{\mathsf{upper}} \bSigma_{\mathsf{upper}}
  \widehat{\bV}_{2}^{\top}$ of $\Mstar_b$. \\
  \tcc{Step 3: Imputation of missing entries} Compute the matrix
  $\Mhat_{d} \coloneqq \widehat{\bU}_{2}( \widehat{\bU}_{1}^{\top}
  \widehat{\bU}_{1})^{-1} \widehat{\bU}_{1}^{\top}
  \Mhat_{b}$. \\
  \textbf{Output:} $\Mhat_{d}$ as estimate of $\Mstar_{d}$, along
  with intermediate quantities $(\bUhat, \bVhat)$.
  \caption{\FourBlockEst \label{alg:4-blocks-Md}}
\end{algorithm}

Recall the partially observed matrix $\ObsMatrix$ defined in
equation~\eqref{Eqn4Block}, and define
\begin{align*}
  \ObsLeft \defn  \left [\begin{array}{c} \ObsMatrix_{a}
    \\ \ObsMatrix_{c}
  \end{array}  \right ] \in \RR^{N\times T_1},  \qquad 
  \ObsUpper \defn  \left [\, \ObsMatrix_{a}\;\; \ObsMatrix_{b}\,
   \right ] \in \RR^{N_1 \times T}
\end{align*}
as its left and upper submatrices.  Algorithm \FourBlockEst~takes as
input the matrix $\ObsMatrix$, and then operates separately on these
submatrices of the observation matrix.  It returns as output a matrix
estimate $\Mhat_d$ of the unknown block $\Mstar_d$ of the matrix
$\Mstar$; see equation~\eqref{Eqn4Block}.

To understand the purpose of the different steps in Algorithm
\FourBlockEst, consider the singular value decomposition (SVD) of the
matrix $\Mstar$---say $\Mstar = \bU^\star \SigStar (\bV^\star)^\top$,
where $\bU^\star \in \real^{N \times r}$ and $\bV^\star \in \real^{T
  \times r}$ are matrices of left and right singular vectors.  The
three main steps of the algorithm are:
\begin{cdesc}
\item[Left subspace estimation:] The matrix $\bUhat$ from Step 3
  is an estimate of $\bU^\star$.
\item[Right subspace estimation:]  Similarly, the matrix $\bVhat$
  in Step 5 is an estimate of $\bV^\star$.
\item[Matrix denoising:] Step 6 uses the appropriate components of
  $(\bUhat, \bVhat)$ to compute a denoised estimate of $\Mstar_b$.
\item[Matrix imputation:] Step 7 combines the denoised estimate
  with the subspace estimate to compute an estimate of $\Mstar_d$.
\end{cdesc}

As discussed in our previous paper~\cite{yan2024entrywise}, in the
idealized case that $\Mstar$ is rank $r$, and we observe the blocks
$\{\Mstar_a, \Mstar_b, \Mstar_c \}$ without noise, then the output
$\Mhat_d$ of Algorithm \FourBlockEst~is guaranteed to be equivalent to
$\Mstar_d$.  The same paper also analyzes its estimation-theoretic
properties in the more realistic setting of noisy observations.


\subsubsection{Confidence intervals for $4$-block designs}
\label{SecConf4Block}

Having specified our estimation procedure (Algorithm \FourBlockEst),
we now describe how to use its outputs for computing confidence
intervals for each entry of $\Mstar$.  This routine, specified
precisely as Algorithm \FourBlockConf, involves three key steps.
  
\begin{algorithm}[h]
  \DontPrintSemicolon \SetNoFillComment
  \textbf{Input:} Confidence level $1-\alpha \in (0,1)$ and inputs of
  Algorithm~\FourBlockEst. \\
  Call Algorithm \FourBlockEst~to compute $(\Mhat_d, \widehat{\bU},
  \widehat{\bV})$.  \\
  \tcp{Residual Estimates} 
  Estimate the residuals $\widehat{\bE}$
  via equation~\eqref{eq:noise-estimator}. \\
  \For{$ i = N_1 + 1$ \KwTo $N$}{ \For{$ t = T_1 + 1$ \KwTo $T$}{
            \tcp{Variance Estimates} 
      Compute the variance estimate $\widehat{\estvar}_{i,t}$ from
      equation~\eqref{eq:gamma-hat}.

      \tcp{Confidence Intervals} 
      Compute the interval $\mathsf{CI}_{i,t}^{(1-\alpha)} \coloneqq
      [\widehat{M}_{i,t} \pm \Phi^{-1} (1 - \alpha/2) \,
        \widehat{\gamma}_{i,t}^{1/2} ]$.
  } } \textbf{Output:} Return $\mathsf{CI}_{i,t}^{(1-\alpha)}$ as
  confidence interval $M_{i,t}^\star$ for each unobserved $(i,t)$.
  \caption{\FourBlockConf\label{alg:4-blocks-Md-CI}}
\end{algorithm}

\begin{subequations}
\begin{cdesc}
\item[Estimating the residuals:] Use the output $\Mhat$ to estimate
  the residuals $\Emat \defn \ObsMatrix - \Mstar$ via
\begin{align}
\label{eq:noise-estimator}
\widehat{\bE} \coloneqq \left [\begin{array}{cc} \bM_{a} - \Mhat_a &
    \bM_{b} - \Mhat_b \\ \bM_{c} - \Mhat_c & \bm{?}
    \end{array}  \right ] \quad \text{where} \quad 
     \left [\begin{array}{c} \Mhat_a \in \RR^{N_1 \times T_1} \\ \Mhat_c
        \in \RR^{N_2 \times T_1}
    \end{array}  \right ] \coloneqq \bU_{\mathsf{left}} \bSigma_{\mathsf{left}} \bV_{\mathsf{left}}^\top.
\end{align}
The bottom right block of $\widehat{\bE}$ is not defined, just as for
$\bE$.
\item[Variance estimation:]
  Partition the subspace estimates $(\bUhat, \bVhat)$ as
  \begin{align}
\label{EqnHatPartition}    
  \hatbU \defn \left [\begin{array}{c} \hatbU_{1} \in \mathbb{R}^{N_1
        \times r} \\ \hatbU_{2} \in \RR^{N_2 \times r}
  \end{array}  \right ],  \qquad
  \hatbV \defn \left [\begin{array}{c} \hatbV_{1} \in \RR^{T_1 \times
        r} \\ \hatbV_2 \in \RR^{T_2 \times r}
  \end{array}  \right ]. 
\end{align}    
Use the estimated residuals $\Ehat$ and $(\bUhat, \bVhat)$ to form the
variance estimate
\begin{align}
\label{eq:gamma-hat}
\widehat{\estvar}_{i,t} \coloneqq \sum_{k=1}^{N_1}
\widehat{E}_{k,t}^{2} \Big[ \hatbU_{i,\cdot}
  \underbrace{(\hatbU_1^{\top} \hatbU_1 )^{-1}}_{\in \real^{r \times
      r}} \hatbU_{k,\cdot}^{\top} \Big]^{2} + \sum_{s=1}^{T_1}
\widehat{E}_{i,s}^{2}  \left [ \hatbV_{t,\cdot} (\hatbV_1^{\top}
  \hatbV_1)^{-1} \hatbV_{s,\cdot}^{\top}  \right ]^{2},
\end{align}
where $\bUhat_{i, \cdot} \in \real^r$ is the $i^{th}$ row of $\bUhat$,
and $\bVhat_{s, \cdot} \in \real^r$ is the $s^{th}$ row of $\bVhat$.
\item[Confidence intervals:] Given a level $\alpha \in (0,1)$, for
  each unobserved entry $(i,t)$, we construct the interval
\begin{align}
  \label{eq:CI-construction}
  \mathsf{CI}_{i,t}^{(1-\alpha)} \coloneqq \Big[\widehat{M}_{i,t} \pm
    \Phi^{-1} \big(1 - \tfrac{\alpha}{2} \big)
    \widehat{\gamma}_{i,t}^{1/2} \Big],
\end{align}
where $\Phi$ denotes the CDF of the standard normal distribution.
\end{cdesc}
\end{subequations}

As discussed in~\Cref{SecCoverage}, under appropriate regularity
conditions, the interval $\mathsf{CI}_{i,t}^{(1-\alpha)}$ is
guaranteed to include the unknown matrix entry $M_{i,t}^\star$ with
probability converging to $1 - \alpha$.  For short, we say that it is
a confidence interval (CI) with coverage $1 - \alpha$.  We also
comment that the variance estimate~\eqref{eq:gamma-hat} is motivated
by calculations of the asymptotic variance of the error in the matrix
estimate $\Mhat_d$ returned by Algorithm \FourBlockEst.  See the
discussion following~\Cref{ThmInformal} for more details.


\subsection{Confidence intervals for general staggered design}
\label{SecStaggered}

With our two sub-routines (namely, Algorithms \FourBlockEst~and
\FourBlockConf) in place, we are now ready to describe our final
inference routine, which applies to the general staggered design.

\begin{algorithm}[h!]
  \DontPrintSemicolon \SetNoFillComment
  \textbf{Input:} Data matrix $\ObsMatrix$, rank $r$, confidence level
  $1-\alpha$. \\
  Extract the dimension information $\{N_i\}_{1 \leq i \leq k}$ and
  $\{T_j\}_{1 \leq j \leq k}$ from $ \bM$. \\
  \For{$ i_0 = 1$ \KwTo $k$}{ \For{$j_0 = k + 2 - i_0$ \KwTo $k$}{
      Construct the four-block data matrix $\InputBlock{i_{0}}{j_{0}}$ via
      equation~\eqref{eq:reduction}. \\
      Call Algorithm \FourBlockEst~with input $\InputBlock{i_{0}}{j_{0}}$
      and rank $r$ to compute an estimate $\Mhat_d$ for its
      bottom-right block.\\
      Extract the appropriate submatrix of $\Mhat_d$, denoted by
      $\MhatBlock{i_0}{j_0}$, as the estimate for the block
      $\MstarBlock{i_0}{j_0}$. \\
      Call Algorithm \FourBlockConf~with input $\MhatBlock{i_0}{j_0}$
      and confidence level $1 - \alpha$ to compute entry-wise
      confidence intervals for $\MstarBlock{i_0}{j_0}$.  } }
  \textbf{Output:} $\widehat{M}_{i,t}$ as the estimate, and
  $\mathsf{CI}_{i,t}^{(1-\alpha)}$ as the confidence interval, for
  each unobserved entry $M_{i,t}^\star$.
  \caption{\StaggeredConf
    \label{alg:general}}
\end{algorithm}

This routine is based on a partitioning scheme, described
in~\Cref{AppPartition}, that takes as input a staggered design,
returns a positive integer $k$, and an associated set of dimensions
$\{N_i\}_{i=1}^k$ and $\{T_j\}_{j=1}^k$.  Via these objects, for any
pair of integers $i_0 \in [k]$ and $j_0 \in (k + 1 - i_0, k]$, we can
  extract an observation matrix $\InputBlock{i_{0}}{j_{0}}$
  corresponding to a particular four-block problem; in particular, see
  equation~\eqref{eq:reduction} in~\Cref{AppPartition} for the
  details.  For each such pair of indices, we first perform matrix
  estimation by applying Algorithm \FourBlockEst~to this sub-problem,
  and then using its outputs, we apply Algorithm \FourBlockConf~to
  compute confidence intervals (CIs) for each of the unobserved
  entries within this sub-problem.  Our construction ensures that each
  unobserved matrix entry in the full staggered matrix is covered by
  one of these four-block sub-problems.  Thus, after running Algorithm
  \StaggeredConf, we have produced a CI for each of the unobserved
  entries associated with the original staggered design.

  
\subsection{Coverage guarantees}
\label{SecCoverage}

Thus far, we have described a routine (Algorithm \StaggeredConf) that,
for any level $\alpha \in (0, 1)$ and unobserved index $(i,t)$,
produces an interval $\mathsf{CI}_{i,t}^{(1-\alpha)}$.  We now turn to
the coverage properties of this interval: in particular, we would like
to ensure that it contains the unknown value $M^\star_{i,t}$ with
probability converging to $1 - \alpha$ as either $N$ or $T$ grow.

In order to provide a guarantee of this type, we need to impose
certain assumptions on the problem, and in particular, on the noise
random variables $\noise_{i,t} \defn Y_{i,t} - M^\star_{i,t}$ indexed
by $i \in [N]$ and $t < t_i$.  We assume that they are independent,
zero-mean sub-Gaussian random variables with variances
$\noisevar_{i,t}^2 \defn \mathsf{var}( \noise_{i, t} )$, and we define
\begin{align}
\label{EqnHetero}  
    \sigmax \defn \max_{ \substack{i \in [N] \\ t < t_i}}
    \sigma_{i,t}.
\end{align}
This assumption and mild variants of it are commonly made in the
literature on panel data (e.g.,~\cite{eli21synthcontstag,
  athey2021matrix, borusyak2024revisiting}).  In addition, as is
standard in the matrix completion literature, we require certain
incoherence conditions on the matrix $\Mstar$; see~\Cref{AppTheory}
for the details.

\begin{theorem}[Informal version, see~\Cref{AppTheory} for precise version]
\label{ThmInformal}
Under certain regularity conditions, for any four-block problem with
rank $r \precsim \min \{N_1, T_1 \}$ and maximal noise level
$\sigmax$, we have
\begin{align}
\label{EqnInformal}  
    \PP\Big[ \mathsf{CI}_{i,t}^{(1-\alpha)} \ni M^\star_{i,t} \Big]
    \geq 1 - \alpha - \mathcal{O} \Big( \frac{\sigmax \: \polylog(N,
      T)}{\sqrt{\min \{N, T \}}} \Big ).
  \end{align}
\end{theorem}
\noindent Here $\polylog(N, T)$ denotes a quantity that depends
polynomially on $\log(N + T)$.  Consequently, the interval
$\mathsf{CI}_{i,t}^{(1-\alpha)}$ is guaranteed to provide the
prescribed coverage as $N$ and $T$ go to infinity.

\paragraph{Some proof intuition:}
While the proof itself is given in~\Cref{AppTheory}, let us provide
some intuition here.  Let the blocks $(\bUstar_1, \bUstar_2)$ and
$(\bVstar_1, \bVstar_2)$ be defined as in the partition of
equation~\eqref{EqnHatPartition}, and define the random matrix
\begin{align}
\label{eq:Z-defn}  
\bZ & \coloneqq \bU_2^\star(\bU_1^{\star\top}\bU_1^\star)^{-1}
\bU_1^{\star\top} \bE_{b} + \bE_{c} \bV_1^\star (\bV_1^{\star\top}
\bV_1^\star)^{-1} \bV_2^{\star\top}.
\end{align}
We begin by showing that our estimate $\Mhat_d$ satisfies an
approximation of the form \mbox{$\Mhat_d \approx \bM_d^\star
  +\bZ$,} where the remainder term can be controlled in an
entrywise sense.  Since $\bZ$ is linear in the independent
mean-zero noise components $\bE_b$ and $\bE_c$, it is can be shown
(e.g., using Berry-Esseen type of arguments) that each entry $Z_{i,t}$
is approximately Gaussian with zero mean and variance
\begin{align}
\label{eq:defn-variance}
  \estvar_{i,t}^\star \defn \mathsf{var} \left ( Z_{i,j} \right ) =
  \sum_{k=1}^{N_1}\noisevar_{k,t}^{2} \left [ \bU_{i,\cdot}^\star
    (\bU_1^{\star\top}\bU_1^\star)^{-1}\bU_{k,\cdot}^{\star\top}
    \right ]^{2} + \sum_{k=1}^{T_1}\noisevar_{i,k}^{2} \left [
    \bV_{t,\cdot}^\star
    (\bV_1^{\star\top}\bV_1^\star)^{-1}\bV_{k,\cdot}^{\star\top}
    \right ]^{2},
\end{align}
Thus, we see the motivation for the variance estimate
$\widehat{\estvar}_{i,t}$ from equation~\eqref{eq:gamma-hat}
in Algorithm \FourBlockConf.  Given this high-level road map,
the bulk of the proof involves establishing that the estimate
$\widehat{\estvar}_{i,t}$ is ``close enough'' to the true
variance~\eqref{eq:defn-variance} so as to guarantee the
claimed coverage guarantees.  See~\Cref{AppTheory} for the details. \\

\paragraph{Confidence intervals for bilinear forms:}
Finally, let us comment on a natural and important extension to
confidence intervals (CIs) for individual entries of the matrix
$\Mstar_d$.  A more general problem is to produce CIs for a bilinear
function of its entries: more precisely, given arbitrary vectors
$\lvec \in \RR^{N_2}$ and $\rvec \in \RR^{T_2}$, suppose that we
want a CI for the scalar $\lvec^\top \bM_d^\star \rvec$.  This
set-up allows us to handle the weighted treatment
effect~\eqref{eqn:WTE} as a special case by setting $\lvec$ to be a
canonical basis vector.  It also allows us to estimate various other
types of \emph{aggregated treatment effects}, in which we weight both
the units and time periods in an arbitrary period (states and years,
for the ACA example that we discuss in detail).
See~\Cref{AppBilinear} for details on how to produce CIs for bilinear
forms.


\subsection{Numerical studies}
\label{sec:simulations}

In this section, we report the results of some numerical studies to
understand the coverage and widths of the confidence intervals
returned by the CAST method.  Moreover, we compare its performance of
with that of a synthetic control method (SCM). Synthetic controls are
some of the most popular methods for analyzing the causal effects of
policy decisions; we make use of the \texttt{augsynth} method that is
applicable to panel data with staggered adoption, and comes equipped
with a software library~\cite{ben2021augmented}, rendering it an
excellent benchmark for comparison.  There are also other
matrix-completion methods~\cite{choi2023matrix, athey2021matrix} for
causal analysis, under staggered adoption, albeit without software
implementation. However, there are various theoretical reasons why our
matrix-based method is preferable; see the discussion
in~\citet{yan2024entrywise} for further details.

We make these comparisons using a semi-synthetic dataset that is
generated from Affordable Care Act data via the procedure described
below. This set-up enables comparisons to a ``ground truth'', so that
the accuracy of different methods can be compared directly.  We
provide comparisons between the CAST and SCM methods for both
estimating the ITE and the ATET; however, it should be noted that the
analysis in the paper~\citep{eli21synthcontstag} only provides
estimation guarantees for the ATET, and does not establish any sort of
asymptotic normality or validity of the proposed confidence intervals.
Despite the absence of such guarantees, it is still interesting to
compare the ITE and ATET estimates along with their standard errors,
as provided by the \texttt{augsynth} package, with the results of the
CAST method.

\subsubsection{Semi-synthetic data}
We make our comparisons for several different choices of $(N, T)$
pairs using semi-synthetic data generated based on the medical
uninsurance rates data from the ACA dataset.  Beginning with this
dataset of uninsurance rates (see~\Cref{SecDescribe} for details), we
use it to estimate a factor model 
\begin{align*}
\mat{Y} = \sum_{j=1}^r \sing_j^\star \bU_{j}^\star (\bV_{j}^\star)^T +
\noise_{it},
\end{align*}
where the vector $\sing^\star \in \real^r$ and matrices
\mbox{$\bU^\star \in \real^{N \times r}$} and \mbox{$\bV^\star \in
  \real^{T \times r}$} are the model parameters.  We choose $r$ based
on the singular values of the observed matrix in the same manner as
when we run the our method, such as the ``hockey stick'' visual
heuristic when looking at the scree plot.

Given estimates $(\singhat, \bUhat, \bVhat)$ of these quantities, we
then form the matrix
\begin{align*}
\bUhat_{\singhat} = \begin{bmatrix} \vert & \vdots & \vert
  \\ \sqrt{\singhat_{1}} \cdot \bUhat_1 & \vdots & \sqrt{\singhat_{r}} \cdot \bUhat_r \\ \vert &
  \vdots & \vert \end{bmatrix},
\end{align*}
with the quantity $\bVhat_{\singhat}$ defined in an analogous
manner. We then let $\widehat{u} \in \real^r$ and $\SigHat_{u} \times
\real^{r \times r}$ be the mean and covariance, respectively, of the
rows of $\bUhat_{\singhat}$, and the same for $\widehat{v}$ and
$\SigHat_v$ with respect to $\bVhat_{\singhat}$. We also compute the
empirical variance
\begin{align*}
\widehat{\sigma}^2_{\noise} \defn \frac{1}{NT} \sum_{i, t} \| \mat{Y}
- \bUhat_{\singhat} \bVhat_{\singhat}^T \|_F^2.
\end{align*}

Given the fitted parameters $(\widehat{u}, \widehat{v}, \SigHat_u,
\SigHat_v, \widehat{\sigma}^2_{\noise})$, we generate a panel dataset
of rank $r$ and dimensions $(N', T')$ according to the following three
steps:
\begin{description}
\item[Step 1:]
Sample the $r$-dimensional random vectors
\begin{align*}
\phi_i &\sim \Normal(\widehat{v}, \SigHat_v), \quad \mbox{for $i = 1,
  \ldots, N'$, and} \quad \mu_t \sim \Normal(\widehat{u}, \SigHat_u)
\quad \mbox{for $t = 1, \ldots, T'$,}
\end{align*}
along with the additive noise variables
\begin{align*}
  \noise_{i,t} &\sim \Normal(0, \widehat{\sigma}^2_{\noise} ) \quad
  \mbox{for all pairs $(i, t) \in [N'] \times [T']$.}
\end{align*}
\item[Step 2:] Generate the responses
\begin{align*}
  Y_{i,t} = \langle \phi_i, \mu_t \rangle + \noise_{i,t} \qquad
  \mbox{for all pairs $(i, t) \in [N'] \times [T']$.}
\end{align*}
\item[Step 3:] Generate a random collection of treatment times
  $\{t_i\}_{i=1}^{N'}$ that are uniformly distributed between
  $[0.7N', 1.3N']$.
\end{description}

\noindent These steps are repeated $1000$ times to generate the
results presented in the next section.

\subsubsection{Comparison of CAST and SCM}

We begin by comparing the mean-squared errors (MSEs) of the CAST
estimates with those of the SCM estimates, computed using the Augsynth
package~\cite{ben2021augmented}.
\begin{figure}[ht!]
\begin{center}
\begin{tabular}{ccc}
  \widgraph{0.45\textwidth}{figures/ITE_MSE} &&
  \widgraph{0.45\textwidth}{figures/ATET_MSE} \\ (a) && (b)
\end{tabular}
\caption{Comparison of the mean-squared error (MSE) of the CAST and
  SCM methods for estimating the ITE and ATET.  Standard errors in
  shown as black error bars.  (a) Bar plots of the MSE for estimating
  the individual treatment effects (ITE), with each pair representing
  a different choice of $(N,T)$ as marked.  In all cases, the CAST
  method performs substantially better than the SCM method.  (b)
  Analogous of the MSE for estimating the ATET. In this case, CAST
  still exhibits lower MSE than the SCM method , although the gap in
  performance is not as large as in the ITE setting.}
  \label{fig:MSEs}
\end{center}
\end{figure}
Panels (a) and (b) of~\Cref{fig:MSEs} show, respectively, the MSEs for
estimating (all the individual treatment effect (ITE) and the average
treatment effect on treated (ATET), as well as their standard
errors. We then studied the actual coverage of the confidence
intervals returned by the CAST procedure at a nominal level of
$1-\alpha = 0.9$, leading to a $95\%$ confidence intervals.  As shown
in~\Cref{fig:coverage}, the CAST method exhibits a mild degree of
undercoverage at smaller sample sizes and reaching the desired
coverage at larger sample sizes.\footnote{We remark that for
computational reasons, these coverage results are for the four-block
design adoption; the results do not change meaningfully when
introducing staggered adoption.}

\begin{figure}
\centering
\begin{tabular}{c}
  \widgraph{0.6\textwidth}{figures/CAST_coverage}
\end{tabular}
\caption{Bar plots of actual coverage of CAST confidence intervals at
  a nominal level of $95\%$ for both the ITE and ATET. In these
  simulation, the CAST intervals exhibit some mild undercoverage.}
\label{fig:coverage}
\end{figure}

In our simulations, we observed that the CAST standard erraors are
typically $2$--$3$ times smaller than those of SCM, and that the SCM
approach tends to exhibit overcoverage.  Given these coverage
discrepancies, in order to make a fair power comparison between the
two methods, we proceeded according the following two step procedure:
\begin{enumerate}
\item[(1)] For any given size $\alpha \in (0,1)$ and any method (CAST
  or SCM), we used the method's standard error to compute a critical
  value that ensures exactly $1-\alpha$ coverage.
\item[(2)] Using this critical value, we then computed the power to
  reject the null when the treatment effect is of size $1\%$.
\end{enumerate}
In~\Cref{fig:power}, we plot the size $\alpha$ versus the power for
both the CAST and SCM methods, and for the ITE and ATET in panels (a)
and (b), respectively.  As can be seen from these plots, the CAST
method has substantially higher power for inference on the ITE, along
with a moderate improvement in power for inference on the ATET.

\begin{figure}
\begin{center}
\begin{tabular}{ccc}
  \widgraph{0.45\textwidth}{figures/ITE_power_curve} &&
  \widgraph{0.45\textwidth}{figures/ATET_power_curve} \\ (a) && (b)
\end{tabular}
\caption{Plots of the size $\alpha$ versus the power of tests,
  comparing with the CAST method with the SCM mehod.  (See main text
  for further details on how the power was computed.) (a) Plots of the
  size $\alpha$ against the power for the ITE.  In this case, the CAST
  approach provides substantially higher power.  (b) Plots of the size
  $\alpha$ against the power for the ATET; here CAST yields more
  moderate improvements in power over SCM.}
  \label{fig:power}
\end{center}
\end{figure}


\section{Case study: Analysis of the Affordable Care Act}
\label{sec:results}

We now return to the Affordable Care Act, first introduced
in~\Cref{sec:ACA}, and apply the proposed procedures for inferring
treatment effects.  At a high level, we apply our method to three
different outcomes:
\begin{itemize}[leftmargin=1em,itemsep=0pt]
\item uninsurance rates
\item healthcare expenditures, and
\item   infant mortality.
\end{itemize}
When applied to any one of these outcomes, the final output of our
procedure is an matrix estimate $\Mhat$ (or equivalently, a collection
of estimates of treatment effects), along with confidence intervals
for these quantities.  Thus, our results can be viewed as a collection
of time series, one for each state.  For states that adopt treatment,
we have an observed set of pre-treatment outcomes, an observed set of
post-treatment outcomes, and our method provides estimates of the
counterfactual outcomes that would have been observed if treatment
\emph{had not} been adopted, along with confidence intervals.
\Cref{fig:results-graphs} plots our results in this time series format
for two different states, namely New York and Montana; see the figure
caption for further details. As detailed in our simulations, we report
results based on $99\%$ confidence intervals.

\begin{figure}[h!]
\begin{center}
\begin{tabular}{ccc}
  \widgraph{0.45\textwidth}{\uninsurdir/NY} &&
  \widgraph{0.45\textwidth}{\uninsurdir/MT} \\ (a) && (b)
  \\ \widgraph{0.45\textwidth}{\healthdir/NY} &&
  \widgraph{0.45\textwidth}{\healthdir/MT} \\ (c) && (d)
  \\ \widgraph{0.45\textwidth}{\infantdir/NY} &&
  \widgraph{0.45\textwidth}{\infantdir/MT} \\ (e) && (f)
  \end{tabular}
  \caption{Results of Medicaid expansion over time for New York (NY)
    and Montana (MT) on uninsurance rates (plots (a) and (b)), health
    expenditures per capita (plots (c) and (d)), and infant
    mortalities per 100,000 births (plots (e) and (f)). Black
    represents values prior to the adoption of the expansion, and blue
    represents the observed measurements after adoption. Red denotes
    the estimated counterfactual of the outcomes under not adopting
    the expansion, and the shaded red denotes the 95\% confidence
    intervals.}
  \label{fig:results-graphs}
\end{center}
\end{figure}

\subsection{Data description}
\label{SecDescribe}
Let us provide some details of the data that underlies our analysis,
and our casting of it within the panel data formulation.  Recall that
we have $N = 50$ states, and a subset of $40$ of them have chosen to
adopt Medicaid expansion at different times; see~\Cref{fig:expansion}
for a graphical illustration.  In most cases, a state that chose to
adopt the expansion in a given year did so in January; accordingly, we
assigned the first treatment period of those states to be the year of
expansion. For other states with implementation in other months, we
selected the year of the closest January as the first treatment
period.

Specifically, we are interested in analyzing the causal effects of
Medicaid expansion on three different outcomes: uninsurance rates,
economics, and infant mortality rates.  Each of these outcomes have
been studied in some past work.  For instance, the 2008 Oregon
Medicaid expansion studies~\cite{finkelstein2012oregon,
  baicker2013oregon} estimated that the effect of Medicaid expansion
reduced uninsurance rates by about 25\% in low-income adults. Some
past work~\cite{miller21medicaid} has reported that Medicaid expansion
had reduced mortality rates, whereas another
paper~\cite{bhatt2018medicaid} reported that infant mortality had
risen in non-Medicaid expansion states and dropped in Medicaid
expansion states between 2014 and 2016.

\paragraph{Uninsurance rates:} Our data for the uninsured rates
is collected by the Kaiser Family Foundation (KFF) which contains
uninsurance rates for each state between 2008 and 2022.  Our goal is
to evaluate whether or not the Medicaid expansion has achieved its
stated goal of reducing uninsurance rates in states that have adopted
the policy.

\paragraph{Healthcare spending:} As with any policy intervention,
the benefits of Medicare expansion need to considered in conjunction
with its costs.  Some of the arguments against the ACA lie within the
economics of the act. One driving factor for its passage was to
address the extremely high cost of care within the United
States. However, in the years following its passage, health care costs
have continued to grow; for this reason, opponents of ACA argue that
Medicaid expansion is ineffective in addressing these issues.  We use
our inferential methods in an an attempt to address this important
question, in particular via the outcome variable defined as the effect
of Medicaid expansion on healthcare spending per capita.  The data is
taken from the Kaiser Family Foundation.

\paragraph{Infant mortality:} As mentioned in the introduction,
being uninsured plays a significant role in an individual's decision
to seek medical care. It is natural to expect that failure to seek
medical care could lead to affect health outcomes and potentially
higher mortality rates.  In order to address this question, we studied
the effects of Medicare expansion on infant mortality rates, where
``infant'' is defined as a baby less than one year in age.  This data
is taken from the CDC Wonder project~\cite{cdcwonder}. Major causes of
infant mortality include birth defects, premature births as well as
other factors; screening and treating these issues often relies on
repeated checkups. Lack of insurance can reduce one's access to the
requisite medical care, leading to increases in infant mortality.

\subsection{Findings}

We now turn to a more detailed discussion of our findings.

\begin{figure}[h!]
  \begin{center}
  \includegraphics[scale=0.35]{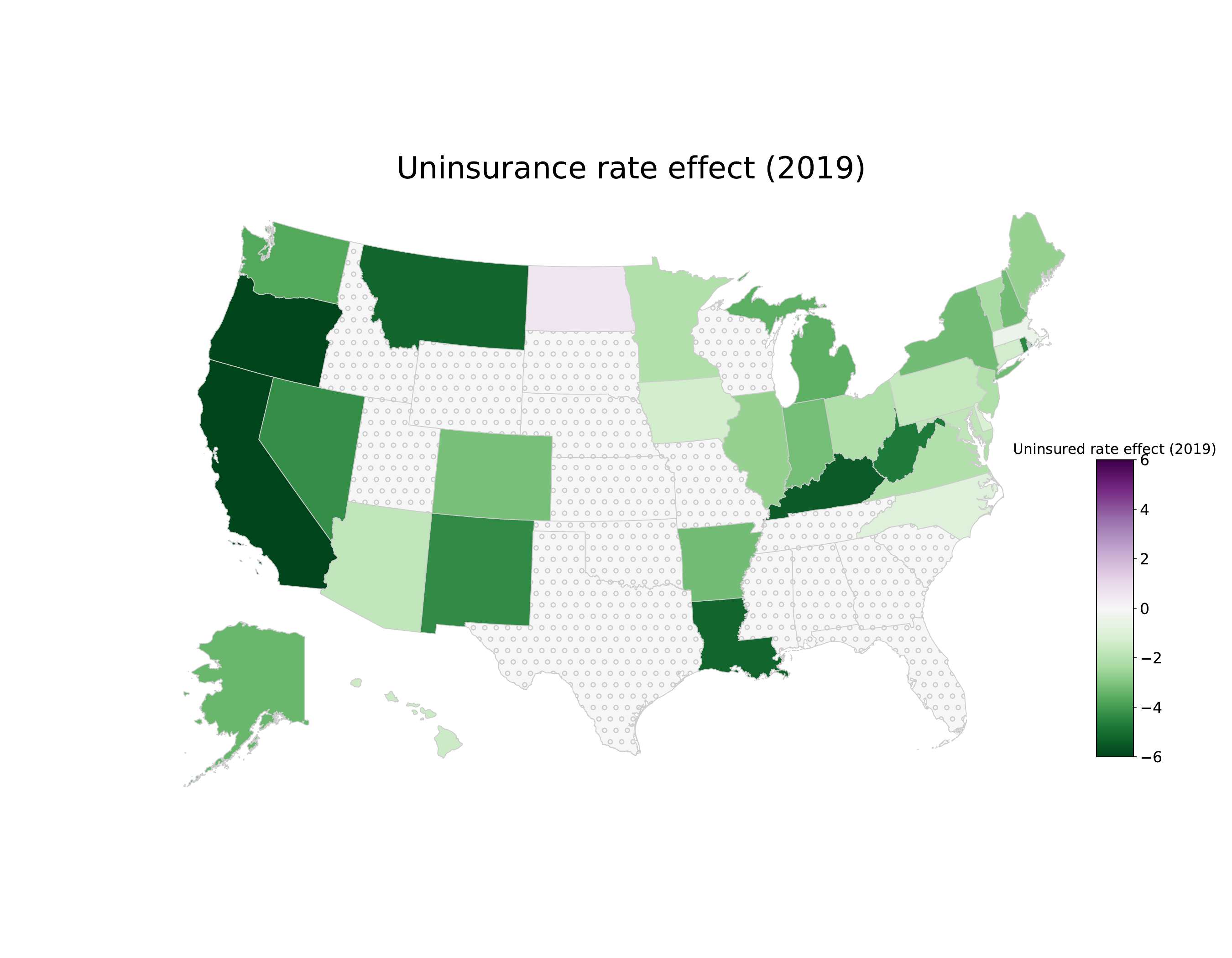}
  \caption{Estimated causal effect of the Medicaid expansion on uninsurance rates in 2019. Measured in terms of absolute percentage decrease. Green (respectively purple) shading indicates a
    decrease (respectively increase) in uninsurance rates. Marked states have not adopted the Medicaid expansion in 2019. As per
    Table~\ref{table:uninsurance}, for this 2019 data, a total of
    0 states reported a significant increase, whereas 29 states
    reported a significant decrease.}
    \label{fig:infant-mortality}
  \end{center}
\end{figure}

\subsubsection{Uninsurance rates}

Table~\ref{table:uninsurance} provides a summary of our results on the
analysis of uninsurance rates.  The first three columns denote the
number of states that have a statistically significant effect in the
specific direction, i.e., in 2022 there were 0 states with a
significant positive effect, 36 states with a significant negative
effect, and 3 states with no significant effects. The fourth column
represents the average effect across states that have adopted
treatment (i.e., the average treatment effect on treated). The final
column represents the population-level effect, i.e., in 2022 we
estimate that approximately 6.5 million Americans are insured as a
result of the Medicaid expansion.  We do not have access to
uninsurance rates in 2020 (likely due to the Covid-19 pandemic) and so
we do not report any results for this year.

\begin{table}[h]
\begin{center}
\caption{Results for uninsurance rates \label{table:uninsurance}}
\begin{tabular}{ l @{\qquad}| c @{ \qquad} c @{\qquad} c @{\qquad} c @{\hspace{0.5em}} c @{\qquad} c @{\hspace{0.5em}}  c }
\hline \hline
 & Positive & Negative & Null & ATET & (SE) & Population Eff. & (SE) \\
 \hline
 2014 & 1 & 23 & 2 & -1.7\% & (0.1) & -3,100,000 & (200,000) \\
 2015 & 1 & 28 & 0 & -2.3\% & (0.1) & -5,000,000 & (100,000) \\
 2016 & 2 & 28 & 1 & -2.4\% & (0.1) & -5,800,000 & (200,000) \\
 2017 & 1 & 28 & 3 & -2.7\% & (0.1) & -6,600,000 & (300,000) \\
 2018 & 1 & 30 & 1 & -2.8\% & (0.1) & -6,800,000 & (100,000) \\
 2019 & 0 & 29 & 5 & -2.6\% & (0.1) & -6,700,000 & (200,000) \\
 2021 & 1 & 36 & 0 & -2.9\% & (0.1) & -7,500,000 & (200,000) \\
 2022 & 0 & 33 & 6 & -2.2\% & (0.2) & -6,500,000 & (400,000) \\
 \hline
\end{tabular}
\end{center}
\end{table}

Based on our analysis, the results in Table~\ref{table:uninsurance}
show that adopting the expansion policy had a substantial effect of
reducing uninsurance rates in the states that have adopted it.
Concretely, for every year between 2014 and 2022, nearly every
adopting state exhibits statistically significant reduction is
uninsurance rates. The estimated average effect from 2015-2022 is
between 2-3\% across the states that have implemented expansion,
resulting in an estimated 6--7 million Americans that are now covered
as a consequence of the Medicaid expansion.  Consequently, based on
our analysis, the Medicaid expansion component of the ACA has
delivered in reducing uninsurance rates in those states that have
adopted it.

\begin{figure}[h!]
  \begin{center}
  \includegraphics[scale=0.35]{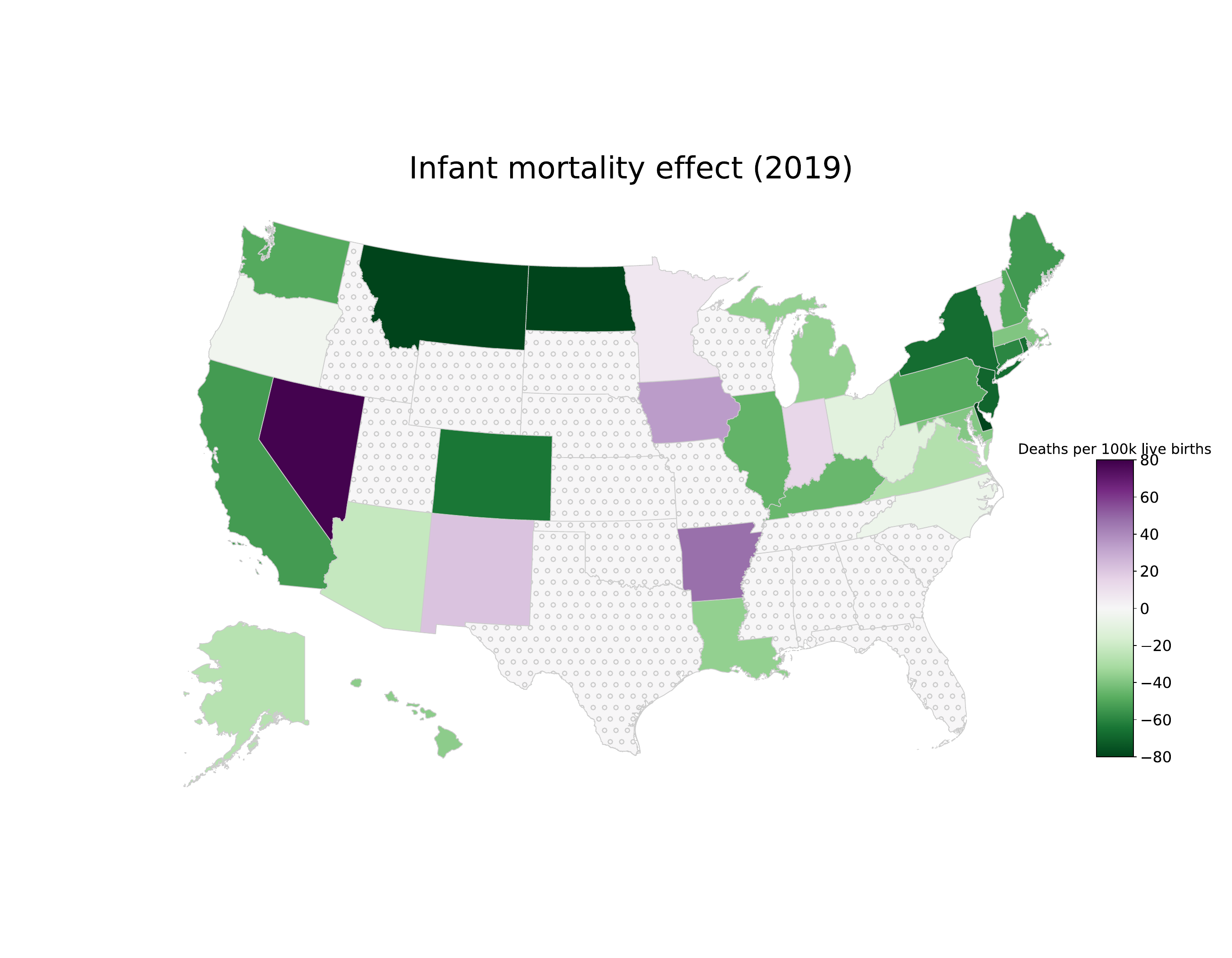}
  \caption{Estimated causal effect of the Medicaid expansion on infant
    mortality rates in 2019. Measured in infant deaths per 100,000
    live births. Green (respectively purple) shading indicates a
    decrease (respectively increase) in infant mortality. Marked states have not adopted the Medicaid expansion in 2019. As per
    Table~\ref{table:infant-mortality}, for this 2019 data, a total of
    3 states reported a significant increase, whereas 22 states
    reported a significant decrease.}
    \label{fig:infant-mortality}
  \end{center}
\end{figure}

\subsubsection{Infant mortality}

We report the results of our analysis in
Table~\ref{table:infant-mortality} and provide a visualization of the
estimated effects in Figure~\ref{fig:infant-mortality}. To clarify,
the measured outcome is actually a moving average with window size $3$
of infant mortalities per 100,000 births. The decision to do so was
made on the basis of the fact that certain states have birth
population adjusted infant mortalities that vary quite dramatically
year to year because within the United States, infant mortalities are
relatively rare events. We can see that Medicaid expansion has
moderate effects on reducing infant mortalities. Every year, there are
around 5 states that have significant increases in newborn deaths, 10
states that have no significant changes, and 15-20 states that have
significant decreases. The average reduction across states between
2014 and 2020 is between 20-30 infant mortalities per 100,000 live
births. For reference, in 2019 across the United States there were 560
reported infant deaths per 100,000 live births. The final column
reports the estimated effect of number of lives saved each year due to
the policy. Our analysis estimates that between 2014 and 2020
approximately 5000 newborns would have died if not for the adoption of
Medicaid expansion.

Figure~\ref{fig:results-graphs} indicates a trend break for infant
mortality: prior to 2014 infant mortality is trending downwards but
our method estimates that infant mortality would have increased under
the counterfactual. This may lead one to doubt the validity of our
causal claims, but we remark that past
studies~\cite{bhatt2018medicaid} have highlighted the fact that infant
mortality rates have, indeed, increased for non-Medicaid expansion
states between 2014-2016. Our method uses the states that did not
adopt Medicaid expansion to impute the counterfactual, so it is quite
reasonable that we would estimate the infant mortality rates to
increase under the control for the states that did adopt the
expansion.

\begin{table}[h]
\begin{center}
\caption{Results of infant mortalities per 100,000 births}
\label{table:infant-mortality}
\begin{tabular}{ l @{\qquad}| c @{ \qquad} c @{\qquad} c  @{\qquad}  c @{\hspace{0.5em}}  c  @{\qquad} c  @{\hspace{0.5em}}  c }
\hline \hline
 & Positive & Negative & Null & ATET & (SE) & Population Eff. & (SE) \\
 \hline
 2014 & 4 & 10 & 12 & -12 &  (3) & -270 & (60) \\
 2015 & 4 & 15 & 9 & -22 & (2) & -640 & (50) \\
 2016 & 4 & 10 & 17 & -26 & (5) & -770 & (100) \\
 2017 & 4 & 14 & 14 & -30 & (3) & -870 & (60) \\
 2018 & 4 & 16 & 12 & -31 & (3) & -880 & (70) \\
 2019 & 3 & 21 & 10 & -33 & (7) & -920 & (140) \\
 2020 & 6 & 18 & 12 & -19 & (5) & -750 & (100) \\
 \hline
\end{tabular}
\end{center}
\end{table}

\subsubsection{Healthcare spending}

Our analysis of the effects of expansion of health expenditures per
capita are reported in Table~\ref{table:health-expenditures}. As
before, the first three columns report the number of states with
statistically significant effects in each direction. We can see from
the table that its unclear whether the expansion caused an increase in
health expenditures per capita.
\begin{table}[h!]
  \begin{center}
\caption{Results of yearly health expenditures per
  capita \label{table:health-expenditures}}
\begin{tabular}{ l @{\qquad}| c @{ \qquad} c @{\qquad} c @{\qquad} c @{\hspace{0.5em}} c }
\hline \hline
 & Positive & Negative & Null & ATET & (SE)  \\
 \hline
 2014 & 7 & 2 & 17 & \$ 99 & (54)  \\
 2015 & 4 & 3 & 22 & \$ 139 & (55)  \\
 2016 & 6 & 5 & 20 & \$ 79 & (57)  \\
 2017 & 5 & 2 & 25 & \$ -92 & (67)  \\
 2018 & 7 & 2 & 23 & \$ 29 & (66) \\
 2019 & 7 & 7 & 20 & \$ -120 & (90)  \\
 2020 & 12 & 6 & 18 & \$ 175 & (65) \\
 2021 & 16 & 8 & 13 & \$ 265 & (49) \\
 2022 & 15 & 10 & 14 & \$ 128 & (61) \\
 \hline
\end{tabular}
\end{center}
\end{table}
In the period 2014---2021, there is a mix of states that have reported
significant increases, decreases, and null effects, on healthcare
spending. Similarly in this vein, the average effect of the policy on
expenditures is quite small, changes of around \$100-200 for most of
the years in that time period, with many of the years reporting
insignificant effects. For reference, in 2019, the health expenditures
per capita across the entire United States was almost \$12,000.
Therefore, our analysis suggests that as opposed to suggestions to ACA
would increase healthcare costs and spending, Medicaid expansion has
negligible effects on healthcare expenditures per capita.


\section{Discussion}
\label{sec:conclusion}

In this paper, we have proposed methodology for estimating and
providing confidence intervals on treatment effects in panel data
under staggered adoption.  Our inferential procedures are sufficiently
flexible to allow for heteroskedastic noise, and to apply to a
generalized notion of treatment effect based on bilinear functionals,
incorporating various types of unit or time-weighted treatment
effects.  In all cases, we provide explicit non-asymptotic coverage
guarantees for the intervals returned by our procedures.

On the methodological front, there are several interesting future
directions to consider.  First, our methodology in this paper, as with
other related work on panel data, relies on a low-rank factor model.
It would be interesting to characterize the robustness of inferential
procedures to departures from the exact form of this factor structure.
Second, in many applications, practitioners have access to additional
covariates that can be used to weaken the identifiability assumptions
that underlie our analysis.  Studying methods of a hybrid type, which
incorporate both factor structure and such additional covariates, is
another important direction for future work.

Finally, we illustrated the utility of our methodology by applying to
to evaluate the effectiveness of Medicaid expansion component of the
the Affordable Care Act (ACA).  We found that Medicaid expansion has
substantially reduced uninsurance rates in the states that have
implemented this policy: concretely, based on our analysis, in 2022,
an estimated 6.5 million Americans now have insurance because of the
adoption of this policy. We also find that the expansion has
negligible effects on increasing healthcare spending.  In terms of
infant mortality, we estimate that between 2014 and 2020, there are
approximately 5,000 newborn babies who would have passed away if not
for Medicaid expansion.  These results contribute to the ongoing
debate on the best approaches to health care in the United States.

\medskip

{\small{
\paragraph{Acknowledgments:}  The authors
thank Alberto Abadie for helpful feedback and suggestions. EX was
funded by an NSF Graduate Research Fellowship; YY was funded by a
Norbert Wiener Postdoctoral Fellowship from the MIT Statistics and
Data Science Center and Institute for Data, Systems, and Society; and
MJW was funded by NSF DMS-2311072 and ONR grant N00014-21-1-2842.}}


\AtNextBibliography{\small}
\printbibliography


\newpage
\appendix

\section{Matrix partitioning for Algorithm \StaggeredConf}
\label{AppPartition}

Any staggered design defines a unique integer $k$ such that $\Mstar$
can be partitioned into a set of $k^2$ blocks $\MstarBlock{i}{j}$ for
$i,j \in [k]$, where the block $\MstarBlock{i}{j} \in \RR^{N_i \times
  T_j}$ is the collection of control outcomes of group $i$ within
stage $j$. Similarly, we partition the full observation matrix
$\ObsMatrix$ so as to obtain submatrices $\ObsBlock{i}{j}$, for $i \in
[k]$ and $j \in [k+1-i]$. The left plot in~\Cref{fig:alg-general}
illustrates this decomposition for a particular staggered design with
$k = 6$.

Given this partition, for each pair of integers $(i_0, j_0)$ such that
$i_0 \in [k]$ and $j_0 \in (k + 1-i_0, \: k]$, our strategy is to
  perform inference for the entries of the block
  $\MstarBlock{i_0}{j_0}$ by reduction to an associated four-block
  problem.  The observation matrix that defines this four-block
  problem is obtained by removing a subset of data. In particular,
  defining the indices $k_{1} =k + 1 - j_{0}$ and $k_{2} = k + 1 -
  i_{0}$, the associated observation matrix $\InputBlock{i_0}{j_0}$
  takes the form 
\begin{align}
  \label{eq:reduction}  
  \InputBlock{i_{0}}{j_{0}} \coloneqq \left [\begin{array}{cc}
      \ObsMatrix_{a} \in \RR^{\bar{N}_1 \times \bar{T}_1} &
      \ObsMatrix_{b} \in \RR^{\bar{N}_1 \times \bar{T}_2} \\
    \ObsMatrix_{c} \in \RR^{\bar{N}_2 \times \bar{T}_1} & \bm{?}
  \end{array}  \right ], \;  \text{where} \;\;
  \begin{cases}
    \ObsMatrix_{a} = \left [ \ObsBlock{i}{j} \right ]_{1 \leq i \leq
      k_{1}, 1 \leq j \leq k_{2}}, \\
    \ObsMatrix_{b} = \left [ \ObsBlock{i}{j} \right ]_{1 \leq i \leq
      k_{1}, k_{2} < j \leq j_{0}}, \\
    \ObsMatrix_{c} = \left [ \ObsBlock{i}{j} \right ]_{k_{1} < i \leq
      i_{0}, 1 \leq j \leq k_{2}},
  \end{cases}
\end{align}
The right plot of~\Cref{fig:alg-general} shows a staggered design with
$k = 6$, and illustrates the construction of four-block problem for
estimating $\MstarBlock{5}{4}$ where the red, blue and yellow blocks
correspond to $\ObsMatrix_a$, $\ObsMatrix_b$ and $\ObsMatrix_c$
respectively.

\section{Coverage guarantees for confidence intervals}
\label{AppTheory}

In this appendix, we state and prove a formal version
of~\Cref{ThmInformal}, our coverage guarantees for the confidence
intervals produced by our procedures.  In addition to the
heteroskedastic noise assumption~\eqref{EqnHetero} stated prior
to~\Cref{ThmInformal}, we impose certain conditions on the expected
outcome matrix $\Mstar$ and its singular vectors, as is standard in
past literature on matrix
completion~\cite{candes2012exact,chen2015incoherence,chen2019inference,yan2021inference}.

\subsection{Precise statement of coverage guarantees}
\label{SecCoveragePrecise}

For a rank $r$ matrix $\Mstar \in \real^{N \times T}$, we have the
singular value decomposition \mbox{$\Mstar = \Ustar \bSigma^\star
  (\Vstar)^\top$,} where $\Ustar \in \real^{N \times r}$ and $\Vstar \in
\real^{T \times r}$ are orthonormal matrices of singular vectors, and
the matrix \mbox{$\bSigma^\star \defn
  \mathsf{diag}\{\sing_1^\star,\ldots,\sing_r^\star\}$} has the
singular values on its diagonal in decreasing order.  For a given
four-block problem with parameters $(N_1, T_1)$, we define $N_2 \defn
N - N_1$ and $T_2 \defn T - T_1$, along with the partitions
\begin{align*}
  \bU^{\star} \defn \left [\begin{array}{c} \bU_{1}^{\star} \in
      \mathbb{R}^{N_1 \times r} \\ \bU_{2}^{\star} \in \RR^{N_2
        \times r}
  \end{array}  \right ],  \qquad
  \bV^{\star} \defn  \left [\begin{array}{c}
    \bV_{1}^{\star} \in \RR^{T_1 \times r} \\ \bV_{2}^{\star}
    \in \RR^{T_2 \times r}
  \end{array}  \right ]. 
\end{align*}
We assume that the top sub-blocks of $\Ustar$ and $\Vstar$ are
\emph{well-conditioned}, in the sense that there exist universal
constants $\clow,\cupper>0$ such that
\begin{subequations}
\begin{align}
\label{EqnSubBlockCondition}    
\clow \frac{N_1}{N} \bI_{r} \preceq \bU_1^{\star \top} \bU_1^\star
\preceq \cupper \frac{N_1}{N}\bI_{r},\qquad \text{and} \qquad \clow
\frac{T_1}{T} \bI_{r} \preceq \bV_1^{\star\top} \bV_1^\star \preceq
\cupper\frac{T_1}{T}\bI_{r}.
\end{align}
Moreover, we assume that they are \emph{incoherent}, meaning that
there is some $\mu \geq 1$ such that
\begin{align}
\label{EqnIncoherent}  
  \left \Vert \bU^\star \right \Vert _{2,\infty} \leq \sqrt{\frac{\mu
      r}{N}}\qquad \text{and} \qquad \left \Vert \bV^\star \right
  \Vert _{2,\infty} \leq \sqrt{\frac{\mu r}{T}}.
\end{align}
\end{subequations}

In terms of the noise variables, we allow them to be
heteroskedastic~\eqref{EqnHetero}, and moreover, we assume that the
sub-Gaussian norm of $\noise_{i,t}$ is of the same order of its
standard deviation, i.e., $ \Vert \noise_{i,t} \Vert _{\psi_2} =
O(\noisevar_{i,t})$.

Our results require certain lower bounds on the signal-to-noise ratio,
or equivalently, upper bounds on the inverse signal-to-noise ratio
(ISNR) given by
\begin{align}
\label{EqnInvSNR}
\SNR & \coloneqq
\frac{\noisevar_{\max}}{\sing_r^\star}\sqrt{\frac{NT}{\min\{N_1,T_1\}}},
\end{align}
where $\noisevar_{\max}$ is the maximum noise
deviation~\eqref{EqnHetero}.  As shown in our past
work~\cite{yan2024entrywise}, any procedure that can consistently
estimate the entries of $\Mstar_d$ requires that $\SNR \to 0$ as the
pair $N$ and $T$ diverge to infinity.

Recall from~\Cref{sec:four-block} that Algorithm
\FourBlockConf~returns, for each unobserved entry $M_{i,t}^\star$, an
interval of the form
\begin{align*}
\mathsf{CI}_{i,t}^{(1-\alpha)} \coloneqq \Big[\widehat{M}_{i,t} \pm
  \Phi^{-1} \big(1 - \tfrac{\alpha}{2} \big) \:
  \widehat{\estvar}_{i,t}^{1/2} \Big],
\end{align*}
where $\widehat{M}_{i,t}$ denotes the $(i,t)$-th entry of the matrix
matrix estimate $\Mhat_d$ returned by Algorithm \FourBlockEst, and the
variance estimate $\widehat{\estvar}_{i,t}$ is defined in
equation~\eqref{eq:gamma-hat}.  The following theorem guarantees the
validity of this confidence interval:
\begin{theorem}
\label{thm:CI}
Consider any target accuracy level $\delta \in (0, 1)$ such that
\mbox{$\min\{N_1,T_1\} \geq \frac{\mu r \log(N + T)}{\delta^2}$,} and
\begin{subequations}
\begin{align}
  \label{eq:signal}
\SNR \leq \frac{\delta}{\sqrt{\mu} r \log^2(N + T)}, \quad \mbox{and}
\quad \SNR \leq \delta \; \max \Big \{ \frac{ \|
  \bU_{i,\cdot}^\star\|_2}{\zeta_{N,T}} \cdot \sqrt{\frac{N}{r}}, \;
\; \frac{ \| \bV_{t, \cdot}^\star\|_2}{\zeta_{N,T}} \cdot
\sqrt{\frac{T}{r}} \Big \},
\end{align}
where $\zeta_{N,T} \coloneqq r \log^{3/2}(N+T) + \sqrt{r}
\log^2(N+T)$.  Then any pair of indices $(i,t)$ corresponding to an
unobserved entry, we have
\begin{align}
\PP \Big [\mathsf{CI}_{i,t}^{(1-\alpha)} \ni M^\star_{i,t} \Big] = 1 -
\alpha + O \big(\delta + \left ( N+T \right )^{-10} \big).
\end{align}
\end{subequations}
\end{theorem}
\noindent See~\Cref{subsec:proof-thm-CI} for the proof. \\


\subsection{Proof of~\Cref{thm:CI}}
\label{subsec:proof-thm-CI}

Our proof makes use of the shorthand notation 
\begin{align*}
\noisevar_{\max} \coloneqq \max_{i,t} \noisevar_{i,t}, \qquad
\noisevar_{\min} \coloneqq \min_{i,t} \noisevar_{i,t}, \qquad \ratio
\coloneqq \frac{\noisevar_{\max}}{\noisevar_{\min}}, \quad \mbox{and} \quad
\spicy \coloneqq \log(N + T).
\end{align*}
We start by presenting a few results that serve as the building block
of our analysis. Recalling the definition~\eqref{eq:Z-defn} of the
matrix $\bZ$, our first lemma shows that the error
$\widehat{M}_{i,t}-M^\star_{i,t}$ can be well-approximated by the
entry $Z_{i,t}$ of $\bZ$.

\begin{lemma}
\label{thm:master}
Under the conditions of~\Cref{thm:CI}, we have the decomposition
\mbox{$\widehat{M}_{i,t}-M_{i,t}^\star = Z_{i,t} + \Delta_{i,t}$,} and
the remainder term $\Delta_{i,t}$ is bounded as
\begin{align*}
  \left | \Delta_{i,t} \right | \leq C_\Delta \; \delta
  \mathsf{var}^{1/2} (Z_{i,j}) = C_\Delta \; \delta
  (\estvar_{i,t}^\star)^{1/2}.
\end{align*}
with probability at least $1 - \smallprob$, for some universal
constant $C_\Delta > 0$.
\end{lemma}
\noindent See~\Cref{subsec:proof-lem-master} for the proof. \\

This lemma demonstrates that the magnitude of $\Delta_{i,t}$ is
dominated by the typical size of $Z_{i,t}$, evaluated by its standard
deviation $\estvar_{i,t}^\star$. Next, we use the Berry–Esseen theorem
to show that $Z_{i,t}$ is approximately Gaussian.
\begin{lemma}
  \label{prop:asym-normality}
Under the conditions of~\Cref{thm:CI}, we have
\begin{align*}
\sup_{s \in \RR} \, \left |\PP \left ( \frac{Z_{i,t} }{
  (\estvar_{i,t}^\star)^{1/2} } \leq s \right ) - \Phi(s) \right | =
O(\delta).
\end{align*}
\end{lemma}
\noindent See~\Cref{subsec:proof-prop-asym-normality} for the proof. \\

Equipped with the above two lemmas, we can show that for any $s \in
\RR$, we have
\begin{align*}
\PP \bigg( \frac{Z_{i,t}+\Delta_{i,t}}{ (\estvar_{i,t}^\star)^{1/2} }
\leq s \bigg) & \leq \PP \bigg( \frac{Z_{i,t}}{
  (\estvar_{i,t}^\star)^{1/2} } \leq s + C_\Delta \delta,\frac{ \left
  |\Delta_{i,t} \right |}{ (\estvar_{i,t}^\star)^{1/2} } \leq C_\Delta
\delta \bigg) + \PP \bigg( \frac{ \left |\Delta_{i,t} \right |}{
  (\estvar_{i,t}^\star)^{1/2} } > \delta \bigg) \nonumber \\
& \overset{\text{(i)}}{\leq}\PP \bigg( \frac{Z_{i,t}}{
  (\estvar_{i,t}^\star)^{1/2} } \leq s + C_\Delta \delta
\bigg)+\smallprob\nonumber \\
& \overset{\text{(ii)}}{\leq}\Phi \left (s + C_\Delta \delta \right )
+ O\big( \delta+ \left ( N+T \right )^{-10} \big)
\overset{\text{(iii)}}{\leq} \Phi(s) + O\big( \delta+ \left ( N+T
\right )^{-10} \big).
\end{align*}
Here step (i) follows from~\Cref{thm:master}, step (ii) utilizes
\Cref{prop:asym-normality}, whereas step (iii) holds since the normal
CDF $\Phi$ is $1/\sqrt{2\pi}$-Lipschitz and under the condition that
$\min\{N_1,T_1\}\geq\delta^{-2}\mu r\spicy$.

Similarly, we have
\begin{align*}
\PP \bigg( \frac{Z_{i,t}+\Delta_{i,t}}{ (\estvar_{i,t}^\star)^{1/2} }
\leq s \bigg) & \geq \PP \bigg( \frac{Z_{i,t}}{
  (\estvar_{i,t}^\star)^{1/2} } \leq s - C_\Delta \delta,\frac{ \left
  |\Delta_{i,t} \right |}{ (\estvar_{i,t}^\star)^{1/2} } \leq C_\Delta
\delta \bigg) \nonumber \\
& \geq \PP \bigg( \frac{Z_{i,t}}{ (\estvar_{i,t}^\star)^{1/2} } \leq s
- C_\Delta \delta \bigg) - \PP \bigg( \frac{ \left |\Delta_{i,t}
  \right |}{ (\estvar_{i,t}^\star)^{1/2} } > C_\Delta \delta
\bigg) \nonumber \\
& \geq \Phi \left (s - C_\Delta \delta \right ) - O\big( \delta+ \left
( N+T \right )^{-10} \big) \geq \Phi(s) - O \big( \delta+ \left ( N+T
\right )^{-10} \big).
\end{align*}
Since the above two relations both hold for any $s \in \RR$, we find
that
\begin{align}
\sup_{s \in \RR} \, \bigg|\PP \bigg( \frac{\widehat{M}_{i,t} -
  M_{i,t}^\star}{ (\estvar_{i,t}^\star)^{1/2} } \leq s \bigg) -
\Phi(s) \bigg | & = \sup_{s \in \RR}\, \bigg| \PP \bigg(
\frac{Z_{i,t}+\Delta_{i,t}}{ (\estvar_{i,t}^\star)^{1/2} } \leq s
\bigg)-\Phi(s) \bigg| \nonumber \\
\label{eq:proof-CI-0}
& = O\big( \delta + \left ( N + T \right)^{-10} \big).
\end{align}

Finally, in order to establish the validity of the data-driven
confidence interval $\mathsf{CI}_{i,t}^{(1-\alpha)}$, we need the
following lemma which shows that $\widehat{\estvar}_{i,t}$ is an
accurate estimate for $\estvar_{i,t}^\star$.

\begin{lemma}
\label{lemma:var-est}
Under the conditions of~\Cref{thm:CI}, we have
\begin{align*}
  \left | \widehat{\estvar}_{i,t} - \estvar_{i,t}^\star \right | \leq
  \frac{\delta}{2\sqrt{\spicy}} \estvar_{i,t}^\star.
\end{align*}
with probability at least $1 - \smallprob$.
\end{lemma}
\noindent See~\Cref{subsec:proof-lemma-var-est} for the proof. \\

Armed with~\Cref{lemma:var-est}, we are now ready to prove the
validity of confidence intervals. We start with the following
factorization
\begin{subequations}
\begin{align}
\label{eq:proof-CI-1}  
 \left | \frac{\widehat{M}_{i,t} - M_{i,t}^\star}{
   (\widehat{\estvar}_{i,t})^{1/2} } - \frac{\widehat{M}_{i,t} -
   M_{i,t}^\star}{ (\estvar_{i,t}^\star)^{1/2} } \right | & = \left
 |\frac{\widehat{M}_{i,t} - M_{i,t}^\star}{
   (\estvar_{i,t}^\star)^{1/2} } \right | \left |
 \frac{\widehat{\estvar}_{i,t} -
   \estvar_{i,t}^\star}{(\widehat{\estvar}_{i,t})^{1/2}
   \big[(\widehat{\estvar}_{i,t})^{1/2} + (\estvar_{i,t}^\star)^{1/2}
     \big]} \right |.
\end{align}
In view of the decomposition $\widehat{M}_{i,t} - M_{i,t}^\star =
Z_{i,t} + \Delta_{i,t}$, we have the following upper bound
\begin{equation}
\label{eq:proof-CI-2}  
\left |\frac{\widehat{M}_{i,t}-M_{i,t}^\star}{
  (\estvar_{i,t}^\star)^{1/2} } \right | \leq \frac{ \left |Z_{i,t}
  \right |}{ (\estvar_{i,t}^\star)^{1/2} }+\frac{ \left |\Delta_{i,t}
  \right |}{ (\estvar_{i,t}^\star)^{1/2} }.
\end{equation}
From the Hoeffding inequality \citep[Theorem
  2.6.3]{vershynin2016high}, we have
\begin{equation}
\label{eq:proof-CI-3}  
\left |Z_{i,t} \right | \leq \widetilde{C}
\sqrt{\estvar_{i,t}^\star\spicy},
\end{equation}
with probability at least $1 - \smallprob$, for some universal
constant. Taken collectively, equations \eqref{eq:proof-CI-2},
\eqref{eq:proof-CI-3} along with~\Cref{thm:master} yield
\begin{equation}
\label{eq:proof-CI-4}  
  \left | \frac{\widehat{M}_{i,t} - M_{i,t}^\star}{
    (\estvar_{i,t}^\star)^{1/2} } \right | \leq
  \widetilde{C}\sqrt{\spicy} + C_\Delta \delta.
\end{equation}
In addition, \Cref{lemma:var-est} guarantees that
$\widehat{\estvar}_{i,t} \geq \estvar_{i,t}^\star/2$, whence
\begin{equation}
\label{eq:proof-CI-4.5}  
  \left | \frac{\widehat{\estvar}_{i,t} -
    \estvar_{i,t}^\star}{(\widehat{\estvar}_{i,t})^{1/2}
    \big[(\widehat{\estvar}_{i,t})^{1/2} + (\estvar_{i,t}^\star)^{1/2}
      \big]} \right |\leq \frac{\delta}{\sqrt{\spicy}}.
\end{equation}
Substituting the bounds \eqref{eq:proof-CI-4} and
\eqref{eq:proof-CI-4.5} into \eqref{eq:proof-CI-1} yields, with
probability at least $1-\smallprob$,
\begin{align}
  \left |\frac{\widehat{M}_{i,t}-M_{i,t}^\star}{
    (\widehat{\estvar}_{i,t})^{1/2}
  }-\frac{\widehat{M}_{i,t}-M_{i,t}^\star}{
    (\estvar_{i,t}^\star)^{1/2} } \right | & \leq
  \big(\widetilde{C}\sqrt{\spicy} + C_\Delta \delta \big)
  \frac{\delta}{\sqrt{\spicy}} \leq 2 C_\Delta
  \delta \label{eq:proof-CI-5}
\end{align}
as long as $\delta \leq 1$ and $C_{\Delta}$ is sufficiently
large. Therefore, for any $s \in \RR$, we have
\begin{align}
\PP \bigg( \frac{\widehat{M}_{i,t}-M_{i,t}^\star}{
  (\widehat{\estvar}_{i,t})^{1/2} } \leq s \bigg) & \leq \PP \bigg(
\frac{\widehat{M}_{i,t}-M_{i,t}^\star}{ (\estvar_{i,t}^\star)^{1/2} }
\leq s + 2 C_\Delta \delta, \bigg |
\frac{\widehat{M}_{i,t}-M_{i,t}^\star}{
  (\widehat{\estvar}_{i,t})^{1/2}
}-\frac{\widehat{M}_{i,t}-M_{i,t}^\star}{ (\estvar_{i,t}^\star)^{1/2}
}\bigg| \leq 2 C_\Delta \delta \bigg)\nonumber \\
& \qquad + \PP \bigg( \bigg|\frac{\widehat{M}_{i,t}-M_{i,t}^\star}{
  (\widehat{\estvar}_{i,t})^{1/2}
}-\frac{\widehat{M}_{i,t}-M_{i,t}^\star}{ (\estvar_{i,t}^\star)^{1/2}
}\bigg| > 2 C_\Delta \delta \bigg) \nonumber\\
& \overset{\text{(i)}}{\leq} \PP \bigg(
\frac{\widehat{M}_{i,t}-M_{i,t}^\star}{ (\estvar_{i,t}^\star)^{1/2} }
\leq s + 2 C_\Delta \delta \bigg)+\smallprob \nonumber \\
& \overset{\text{(ii)}}{\leq}\Phi \left ( s + 2 C_\Delta \delta \right
)+O(\delta+(N+T)^{-10}) \nonumber \\
& \overset{\text{(iii)}}{\leq}\Phi \left ( s \right
)+O(\delta+(N+T)^{-10}). \label{eq:proof-CI-6}
\end{align}
Here step (i) follows from equation~\eqref{eq:proof-CI-5}; step (ii)
follows from equation~\eqref{eq:proof-CI-0}; and step (iii) follows
from the fact that the normal CDF function $\Phi$ is
$1/\sqrt{2\pi}$-Lipschitz.

Similarly, we can show that
\begin{align}
\label{eq:proof-CI-7}  
\PP \bigg( \frac{\widehat{M}_{i,t}-M_{i,t}^\star}{
  (\widehat{\estvar}_{i,t})^{1/2} } \leq s \bigg) \geq \Phi \left ( s
\right )- O (\delta+(N+T)^{-10}).
\end{align}
Taken together, the bounds~\eqref{eq:proof-CI-6}
and~\eqref{eq:proof-CI-7} imply that
\begin{equation}
\label{eq:proof-CI-8}  
\PP \bigg( \frac{\widehat{M}_{i,t}-M_{i,t}^\star}{
  (\widehat{\estvar}_{i,t})^{1/2} } \leq s \bigg) = \Phi \left ( s
\right ) + O(\delta+(N+T)^{-10}) \qquad \mbox{for any $s \in \RR$.}
\end{equation}
\end{subequations}
We conclude that
\begin{align*}
  \PP \big( M_{i,t}^\star \in \mathsf{CI}_{i,t}^{(1-\alpha)} \big) & =
  \PP \bigg( \Phi^{-1} ( \alpha/2 ) \leq
  \frac{\widehat{M}_{i,t}-M_{i,t}^\star}{\sqrt{\estvar_{i,t}}} \leq
  \Phi^{-1} ( 1 - \alpha/2 ) \bigg) \\
  & = \PP \bigg(
  \frac{\widehat{M}_{i,t}-M_{i,t}^\star}{\sqrt{\estvar_{i,t}}} \leq
  \Phi^{-1} ( 1 - \alpha/2 ) \bigg) - \PP \bigg(
  \frac{\widehat{M}_{i,t}-M_{i,t}^\star}{\sqrt{\estvar_{i,t}}} <
  \Phi^{-1} ( \alpha/2 ) \bigg) \\
& \overset{\text{(a)}}{=} 1-\alpha + O(\delta+(N+T)^{-10})
\end{align*}
as claimed, where step (a) follows from equation~\eqref{eq:proof-CI-8}
and its consequence
\begin{align*}
\PP \bigg(
\frac{\widehat{M}_{i,t}-M_{i,t}^\star}{\sqrt{\estvar_{i,t}}} <
\Phi^{-1} ( \alpha/2 ) \bigg) &= \lim_{\varepsilon \to 0+} \PP \bigg(
\frac{\widehat{M}_{i,t}-M_{i,t}^\star}{\sqrt{\estvar_{i,t}}} \leq
\Phi^{-1} ( \alpha/2 ) -\varepsilon \bigg) \\ & = \lim_{\varepsilon
  \to 0+} \Phi \left ( \Phi^{-1} ( 1 - \alpha/2 ) + \varepsilon \right
) + O(\delta+(N+T)^{-10}) \\ & = \frac{\alpha}{2} +
O(\delta+(N+T)^{-10}).
\end{align*}


\subsection{Proof of technical lemmas}

This section provides the proof of
\Cref{thm:master,prop:asym-normality,lemma:var-est}. Before we start,
the following concentration bounds will be useful in the analysis.

\begin{lemma}
  \label{lemma:subgaussian-spectral}
  Let $\bE  \in  \RR^{n_{1}\times n_{2}}$ be a random matrix with
  independent, mean zero, sub-Gaussian entries, and $ \Vert 
  E_{i,j} \Vert _{\psi_{2}} \leq \sigma$.  Then there exists some
  sufficiently large constant $C_g>0$ such that
  \begin{align}
     \left   \Vert  \bE  \right   \Vert  \leq C_g \sigma \sqrt{n}
    \qquad \mbox{where $n \mydefn \max \{ n_{1}, n_{2}
      \}$}
  \end{align}
  holds with probability at least $1 - O(n^{-10})$.  For any $ \bX
   \in  \RR^{n_{1}\times m_{1}}$ and $\bY  \in  \RR^{n_{2}\times
    m_{2}}$,
  \begin{align}
     \Vert  \bX^{\top} \bE \bY  \Vert  \leq \frac{1}{2}
    C_g \sigma  \left   \Vert  \bX  \right   \Vert   \left   \Vert 
    \bY  \right   \Vert  \sqrt{\mathsf{rank}  \left  ( \bX
       \right  ) + \mathsf{rank}  \left ( \bY  \right  ) +
      \log n}
  \end{align}
  holds with probability at least $1 - O(n^{-10})$.
\end{lemma}
\noindent See~\citet[Lemma 19]{yan2024entrywise}.

\subsubsection{Proof of~\Cref{thm:master}}
\label{subsec:proof-lem-master}

We first record a result adapted from \citet[Lemma
  1]{yan2024entrywise}. Define
\begin{align*}
  b_1  \left ( i,t  \right ) & \coloneqq
  \bigg(\frac{\noisevar_{\max}^{3}NT/N_1}{\sing_r^{\star2}\sqrt{N_1/N}}+\frac{\noisevar_{\max}^{2}\sqrt{NT/T_1}}{\sing_r^\star\sqrt{N_1/N}}\bigg) \left  \Vert 
  \bU_{i,\cdot}^\star \right  \Vert  _2\sqrt{r+\spicy},\\ b_2  \left ( i,t
   \right ) & \coloneqq
  \bigg(\frac{\noisevar_{\max}^{3}NT/N_1}{\sing_r^{\star2}\sqrt{T_1/T}}+\frac{\noisevar_{\max}^{2}\sqrt{T+NT/T_1}}{\sing_r^\star\sqrt{T_1/T}}\bigg) \left  \Vert 
  \bV_{j,\cdot}^\star \right  \Vert  _2\sqrt{r+\spicy},\\ b_3  \left ( i,t
   \right ) & \coloneqq \bigg(\SNR^2 \sing_r^\star+
  \noisevar_{\max}\sqrt{\frac{Nr}{N_1}}+\noisevar_{\max}\sqrt{\frac{NT\spicy}{N_1T_1}}\bigg) \left  \Vert 
  \bU_{i,\cdot}^\star \right  \Vert  _2 \left  \Vert 
  \bV_{j,\cdot}^\star \right  \Vert  _2,\\ b_4  \left ( i,t  \right ) &
  \coloneqq \bigg( \frac{\noisevar_{\max}^{3} T}{\sing_{r}^{\star2} }
  \sqrt{\frac{N}{N_1 T_1}} + \frac{\noisevar_{\max}^{3} \sqrt{N_1} }{
    \sing_{r}^{\star2} }\frac{NT}{N_1 T_1} +
  \frac{\noisevar_{\max}^{4}}{\sing_{r}^{\star3}} \sqrt{ \frac{NT}{N_1
      T_1} } \frac{NT}{N_1} \bigg )  \left (r + \spicy  \right  ).
\end{align*}
Then we have the decomposition \mbox{$\widehat{M}_{i,t} -
  M_{i,t}^\star = Z_{i,t} + Z_{i,t}' + \Delta_{i,t}'$,} where
\begin{align*}
Z_{i,t}' \defn (\bE_{c})_{i,\cdot} \bV_1^\star(\bV_1^{\star\top}
\bV_1^\star)^{-1} (\bSigma^\star)^{-1} (\bU_1^{\star\top}
\bU_1^\star)^{-1} \bU_1^{\star\top} (\bE_{b})_{\cdot,t}
\end{align*}
and with probability at least $1-\smallprob$, $| \Delta_{i,j}' | \leq
C_0' \sum_{k=1}^{4} b_{k} ( i,t )$ for some constant $C_0'>0$. It is
worth mentioning that, although~\citet{yan2024entrywise} assumed the
noise are i.i.d.~Gaussian $\mathcal{N}(0,\omega^2)$, Lemma 1 therein
still holds under the heteroskedastic noise
assumption~\eqref{EqnHetero} by replacing $\omega$ with the maximum
noise level $\noisevar_{\max}$.  In addition, we can show that with
probability at least $1-\smallprob$,
\begin{align}
|Z_{i,t}'| & \overset{\text{(i)}}{\leq} C_g \noisevar_{\max} \big \Vert 
\bV_{1}^\star( \bV_{1}^{\star\top} \bV_{1}^\star)^{-1}(
\bSigma^\star)^{-1}( \bU_{1}^{\star\top} \bU_{1}^\star)^{-1}
\bU_{1}^{\star\top}( \bE_{b})_{\cdot,t} \big \Vert _{2} \sqrt{\spicy}
\nonumber \\
& \overset{\text{(ii)}}{\leq} C_g^2 \noisevar_{\max}^2 \big  \Vert 
\bV_{1}^\star( \bV_{1}^{\star\top} \bV_{1}^\star)^{-1}(
\bSigma^\star)^{-1}( \bU_{1}^{\star\top} \bU_{1}^\star)^{-1}
\bU_{1}^{\star\top} \big \Vert  \sqrt{  \left  (r + \spicy  \right  )\spicy}
\nonumber \\
\label{eq:Z'-bound}        
& \overset{\text{(iii)}}{\leq} \frac{C_g^2}{\clow^2}
\frac{\noisevar_{\max}^{2}}{\sing_{r}^\star} \sqrt{\frac{ NT }{ N_1
    T_1}  \left  (r + \spicy  \right  ) \spicy}
\end{align}
Here step (i) applies \Cref{lemma:subgaussian-spectral} conditional on
$\bE_b$; step (ii) utilizes \Cref{lemma:subgaussian-spectral}; step
(iii) follows from condition~\eqref{EqnSubBlockCondition}.

Since the random matrices $\bE_{b}$ and $\bE_{c}$ are independent, we
know that $\mathsf{var}(Z_{i,t}) = \estvar_{i,t}^\star = \alpha_{i,t}
+ \beta_{i,t}$ where
\begin{align*}
  \alpha_{i,t} \coloneqq \mathsf{var}
  \big(\bU_{i,\cdot}^\star(\bU_1^{\star\top}\bU_1^\star)^{-1}\bU_1^{\star\top}(\bE_{b})_{\cdot,t}
  \big) \quad \text{and} \quad \beta_{i,t} \coloneqq \mathsf{var}
  \big((\bE_{c})_{i,\cdot} \bV_1^\star (\bV_1^{\star\top}
  \bV_1^\star)^{-1} \bV_{t,\cdot}^{\star\top} \big).
\end{align*}
In view of the matrix sub-conditioning~\eqref{EqnSubBlockCondition},
we can verify that
\begin{subequations}
  \label{eq:var-lb}
\begin{align}
	\alpha_{i,t} & =\sum_{k=1}^{N_1}\noisevar_{k,t}^{2}\big[
          \bU_{i,\cdot}^\star(\bU_1^{\star\top}\bU_1^\star)^{-1}\bU_{k,\cdot}^{\star\top}
          \big]^{2}
         \in \Big[\cupper^{-1}\frac{N}{N_1}\noisevar_{\min}^{2} \left  \Vert 
          \bU_{i,\cdot}^\star \right  \Vert 
          _2^{2},\clow^{-1}\frac{N}{N_1}\noisevar_{\max}^{2} \left  \Vert 
          \bU_{i,\cdot}^\star \right  \Vert  _2^{2}\Big],\\ \beta_{i,t} &
        =\sum_{k=1}^{T_1}\noisevar_{i,k}^{2}\big[
          \bV_{t,\cdot}^\star(\bV_1^{\star\top}\bV_1^\star)^{-1}\bV_{k,\cdot}^{\star\top}
          \big]^{2}
         \in \Big[\cupper^{-1}\frac{T}{T_1}\noisevar_{\min}^{2} \left  \Vert 
          \bV_{t,\cdot}^\star \right  \Vert 
          _2^{2},\clow^{-1}\frac{T}{T_1}\noisevar_{\max}^{2} \left  \Vert 
          \bV_{t,\cdot}^\star \right  \Vert  _2^{2}\Big].
\end{align}
\end{subequations}
It is straightforward to check that,
\begin{align*}
	\max\Big\{\frac{b_1( i, t )}{ \sqrt{\alpha_{i,t}} },\frac{b_2(
          i, t ) }{ \sqrt{\beta_{i,t}} }\Big\} &
        \overset{\text{(i)}}{\leq} \ratio \cupper^{1/2}
        \sqrt{r+\spicy}(2\SNR+\SNR^2) \overset{\text{(ii)}}{\leq}
        \delta, \\
\frac{b_3( i,t )}{ \sqrt{\alpha_{i,t}+\beta_{i,t}} } &
\overset{\text{(iii)}}{\leq} \ratio \cupper^{1/2} \sqrt{\mu r} \SNR +
\ratio \cupper^{1/2} \sqrt{\frac{\mu r\spicy}{\max\{N_1,T_1\}}} +
\ratio \cupper^{1/2} \sqrt{\frac{\mu r^2}{T}}
\overset{\text{(iv)}}{\leq} \delta,
\end{align*}
where steps (i) and (iii) follow from \eqref{eq:var-lb}, while steps
(ii) and (iv) hold under the conditions of \Cref{thm:CI}. In addition,
\begin{align*}
    b_4' (i,t) + |Z_{i,t}'| & \overset{\text{(a)}}{\leq} 3 \SNR
    \frac{\noisevar_{\max}^{2}}{\sing_r^\star}\sqrt{\frac{NT}{N_1T_1}}
     \left ( r+\spicy  \right ) + \frac{C_g^2}{\clow^2}
    \frac{\noisevar_{\max}^{2}}{\sing_{r}^\star} \sqrt{\frac{ NT }{
        N_1 T_1}  \left  (r + \spicy  \right  ) \spicy}
    \overset{\text{(b)}}{\leq} \delta \estvar_{i,t}^\star,
\end{align*}
where step (a) follows from equation~\eqref{eq:var-lb} and step (b)
holds under the conditions of~\Cref{thm:CI}.  Therefore, we conclude
that with probability at least $1 - \smallprob$, there exists some
universal constant $C_\Delta > 0$ such that
\begin{equation}
 \label{eq:proof-normality-5}  
   \left | \Delta_{i,t}  \right | \leq C_\Delta \delta \mathsf{var}^{1/2}
  (Z_{i,j}) = C_\Delta \delta (\estvar_{i,t}^\star)^{1/2}.
\end{equation}

\subsubsection{Proof of~\Cref{prop:asym-normality}
\label{subsec:proof-prop-asym-normality}}

By definition, we can write
\begin{align*}
Z_{i,t} = \sum_{k=1}^{N_1} \underbrace{\bU_{i,\cdot}^\star
  (\bU_1^{\star\top} \bU_1^\star)^{-1}\bU_{k,\cdot}^{\star\top}
  E_{k,t}}_{\eqqcolon X_{k}} + \sum_{k=1}^{T_1} \underbrace{E_{i,k}
  \bV_{k,\cdot}^\star (\bV_1^{\star\top} \bV_1^\star)^{-1}
  \bV_{t,\cdot}^{\star \top}}_{\eqqcolon Y_{k}}.
\end{align*}
Since the random variables $\{X_{k}:k \in [N_1]\}$ and
$\{Y_{k}:k \in [T_1]\}$ are independent and mean-zero, by the
Berry-Esseen Theorem (see e.g., \citet[Theorem 3.7]{chen2010normal}),
we have
\begin{subequations}
\begin{equation}
\label{eq:proof-normality-1}  
\sup_{s  \in  \RR}\, \left |\PP \left ( \frac{Z_{i,t}}{
  (\estvar_{i,t}^\star)^{1/2} } \leq s \right )-\Phi(s) \right | \leq
\frac{10}{(\estvar_{i,t}^\star)^{3/2}} \left [ \sum_{k=1}^{N_1}\EE
  \big[|X_{k}|^{3}\big]+\sum_{k=1}^{T_1}\EE \big[|Y_{k}|^{3}\big]
  \right ].
\end{equation}
Notice that
\begin{align}
\sum_{k=1}^{N_1}\EE \big[|X_{k}|^{3}\big] & \overset{\text{(i)}}{\leq}
\tilde{c} \sum_{k=1}^{N_1} \left
|\bU_{i,\cdot}^\star(\bU_1^{\star\top}\bU_1^\star)^{-1}\bU_{k,\cdot}^{\star\top}
\right |^{3}\noisevar_{k,t}^{3}\nonumber \\
& \leq \tilde{c} \noisevar_{\max}\max_{k  \in  [N_1]} \big|
\bU_{i,\cdot}^\star (\bU_1^{\star\top} \bU_1^\star)^{-1}
\bU_{k,\cdot}^{\star\top}\big|\sum_{k=1}^{N_1} \noisevar_{k,t}^{2}
\big[\bU_{i,\cdot}^\star(\bU_1^{\star \top}
  \bU_1^\star)^{-1}\bU_{k,\cdot}^{\star \top}\big]^{2} \nonumber \\
& = \tilde{c} \noisevar_{\max}\max_{k  \in  [N_1]} \big|
\bU_{i,\cdot}^\star(\bU_1^{\star\top}\bU_1^\star)^{-1}\bU_{k,\cdot}^{\star
  \top}\big| \alpha_{i,t} \overset{\text{(ii)}}{\leq}
\frac{\tilde{c}}{\clow} \sqrt{\frac{\mu r}{N_1}} \noisevar_{\max}
\left \Vert \bU_{i,\cdot}^\star \right \Vert _2
\alpha_{i,t}. \nonumber \\
\label{eq:proof-normality-2}
& \overset{\text{(ii)}}{\leq} \frac{\tilde{c}}{\clow} \sqrt{ \frac{\mu
    r}{N_1}} \noisevar_{\max} \left \Vert \bU_{i,\cdot}^\star \right
\Vert _2 \alpha_{i,t}.
\end{align}
Here step (i) utilizes the property of sub-Gaussian distribution (see
e.g., \citet[Section 2.5.1]{vershynin2016high}) and $\tilde{c}>0$ is a
universal constant; whereas step (ii) follows from
\begin{align}
 \max_{k \in [N_1]} \big|
 \bU_{i,\cdot}^\star(\bU_1^{\star\top}\bU_1^\star)^{-1}\bU_{k,\cdot}^{\star\top}
 \big| & \leq \left \Vert \bU_{i,\cdot}^\star \right \Vert _2 \big
 \Vert (\bU_1^{\star\top}\bU_1^\star)^{-1}\big \Vert \left \Vert
 \bU_1^\star \right \Vert _{2, \in fty}\nonumber \\ &
 \overset{\text{(a)}}{\leq} \clow^{-1}\sqrt{\frac{N}{N_1}} \left \Vert
 \bU_{i,\cdot}^\star \right \Vert _2\sqrt{\frac{\mu r}{N}} \leq
 \clow^{-1}\sqrt{\frac{\mu r}{N_1}} \left \Vert \bU_{i,\cdot}^\star
 \right \Vert _2,\label{eq:proof-U}
\end{align}
where step (a) utilizes the incoherence
condition~\eqref{EqnIncoherent}.  Using
equation~\eqref{eq:proof-normality-2}, we can show that
\begin{equation}
\label{eq:proof-normality-3}  
\sum_{k=1}^{T_1}\EE \big[|Y_{k}|^{3}\big] \leq \frac{\tilde{c}}{\clow}
\sqrt{\frac{\mu r}{T_1}} \noisevar_{\max} \left \Vert
\bV_{t,\cdot}^\star \right \Vert _2 \beta_{i,t}.
\end{equation}
Combining equations~\eqref{eq:proof-normality-1},
~\eqref{eq:proof-normality-2} and~\eqref{eq:proof-normality-3} yields
\begin{align}
\sup_{s  \in  \RR} \, \left |\PP \left ( \frac{Z_{i,t}}{
  (\estvar_{i,t}^\star)^{1/2} } \leq s \right )-\Phi(s) \right | &
\leq \frac{10 (\tilde{c} / \clow) \noisevar_{\max}
}{(\estvar_{i,t}^\star)^{3/2}} \left [ \sqrt{\frac{\mu r}{N_1}} \left
  \Vert \bU_{i,\cdot}^\star \right \Vert _2
  \alpha_{i,t}+\sqrt{\frac{\mu r}{T_1}} \left \Vert
  \bV_{t,\cdot}^\star \right \Vert _2 \beta_{i,t} \right ]\nonumber
\\ & \overset{\text{(i)}}{\leq} \frac{10 (\tilde{c} /
  \clow)}{(\estvar_{i,t}^\star)^{1/2}} \left (
\noisevar_{\max}\sqrt{\frac{\mu r}{N_1}} \left \Vert
\bU_{i,\cdot}^\star \right \Vert _2+\noisevar_{\max}\sqrt{\frac{\mu
    r}{T_1}} \left \Vert \bV_{t,\cdot}^\star \right \Vert _2 \right
)\nonumber \\ & \overset{\text{(ii)}}{\leq} \frac{10 \ratio \tilde{c}
  \sqrt{\cupper}}{\clow} \sqrt{\frac{\mu r}{\min\{N,T\}}}
\overset{\text{(iii)}}{\leq} \delta \label{eq:proof-normality-4}
\end{align}
\end{subequations}
Here step (i) follows from the fact that
$\estvar_{i,t}^\star=\alpha_{i,t}+\beta_{i,t}$; step (ii) utilizes
\eqref{eq:var-lb}; whereas step (iii) follows from the conditions
given in~\Cref{thm:CI}.


\subsubsection{Proof of~\Cref{lemma:var-est}}
\label{subsec:proof-lemma-var-est}

We first decompose $\estvar_{i,j}^\star = u_{i,t}^\star +
v_{i,t}^\star$ where
\begin{align*}
u_{i,t}^\star \coloneqq \sum_{k=1}^{N_1} \noisevar_{k,t}^{2} \big[
  \bU_{i,\cdot}^\star(\bU_1^{\star\top}\bU_1^\star)^{-1}\bU_{k,\cdot}^{\star\top}
  \big]^{2} \qquad \text{and} \qquad v_{i,t}^\star \coloneqq
\sum_{k=1}^{T_1}\noisevar_{i,k}^{2}\big[
  \bV_{t,\cdot}^\star(\bV_1^{\star\top}\bV_1^\star)^{-1}\bV_{k,\cdot}^{\star\top}
  \big]^{2},
\end{align*}
and $\widehat{\estvar}_{i,t} = \widehat{u}_{i,t} + \widehat{v}_{i,t}$
where
\begin{align*}
\widehat{u}_{i,t} \coloneqq
\sum_{k=1}^{N_1}\widehat{E}_{k,t}^{2}\big[\widehat{\bU}_{i,\cdot}(\widehat{\bU}_1^{\top}\widehat{\bU}_1)^{-1}\widehat{\bU}_{k,\cdot}^{\top}\big]^{2}
\qquad \text{and} \qquad \widehat{v}_{i,t} \coloneqq
\sum_{k=1}^{T_1}\widehat{E}_{i,k}^{2}\big[\widehat{\bV}_{t,\cdot}(\widehat{\bV}_1^{\top}\widehat{\bV}_1)^{-1}\widehat{\bV}_{k,\cdot}^{\top}\big]^{2}.
\end{align*}
In what follows, we will focus on bounding $| \widehat{u}_{i,t} -
u_{i,t}^\star |$; the bound on $| \widehat{v}_{i,t} - v_{i,t}^\star |$
can be established similarly.  To facilitate analysis, we introduce an
intermediate quantity $\overline{u}_{i,t}$ and apply the triangle
inequality to reach
\begin{align*}
  | \widehat{u}_{i,t} - u_{i,t}^\star | \leq | \widehat{u}_{i,t} -
  \overline{u}_{i,t} | + | u_{i,t}^\star - \overline{u}_{i,t} | \qquad
  \text{where} \qquad \overline{u}_{i,t} \coloneqq
  \sum_{k=1}^{N_1}\widehat{E}_{k,t}^{2}\big[
    \bU_{i,\cdot}^\star(\bU_1^{\star\top} \bU_1^\star)^{-1}
    \bU_{k,\cdot}^{\star\top} \big]^{2}
\end{align*}
Recall that $\widehat{E}_{i,t}$ is an estimate for the $(i,t)$-th
entry of the noise matrix.  It takes the form $\widehat{E}_{i,t}^{2} =
\big(M_{i,t} - \widehat{M}_{i,t}\big)^{2} = \big(E_{i,t} +
M_{i,t}^\star - \widehat{M}_{i,t}\big)^{2}\equiv E_{i,t}^{2} +
\delta_{i,t}$, where
\begin{equation}
\label{eq:defn-delta}
\delta_{i,t} \coloneqq \big(M_{i,t}^\star - \widehat{M}_{i,t}\big)^{2}
+ 2 E_{i,t} \big(M_{i,t}^\star - \widehat{M}_{i,t}\big).
\end{equation}
We first study $| u_{i,t}^\star - \overline{u}_{i,t} | $, which can be
further decomposed into two terms
\begin{align*}
  \overline{u}_{i,t}-u_{i,t}^\star &
  =\sum_{k=1}^{N_1}\widehat{E}_{k,t}^{2}\big[
    \bU_{i,\cdot}^\star(\bU_1^{\star\top}\bU_1^\star)^{-1}\bU_{k,\cdot}^{\star\top}
    \big]^{2}-\sum_{k=1}^{N_1}\noisevar_{k,t}^{2}\big[
    \bU_{i,\cdot}^\star(\bU_1^{\star\top}\bU_1^\star)^{-1}\bU_{k,\cdot}^{\star\top}
    \big]^{2}\\ & =\underbrace{\sum_{k=1}^{N_1} \left (
    E_{k,t}^{2}-\noisevar_{k,t}^{2}  \right )\big[
      \bU_{i,\cdot}^\star(\bU_1^{\star\top}\bU_1^\star)^{-1}\bU_{k,\cdot}^{\star\top}
      \big]^{2}}_{\eqqcolon\xi_1}+\underbrace{\sum_{k=1}^{N_1}\delta_{k,t}\big[
      \bU_{i,\cdot}^\star(\bU_1^{\star\top}\bU_1^\star)^{-1}\bU_{k,\cdot}^{\star\top}
      \big]^{2}}_{\eqqcolon\xi_2}.
  \end{align*}
Notice that $\xi_1$ is the sum of $N_1$ independent random variables
\begin{align*}
\xi_1=\sum_{k=1}^{N_1} Z_k \qquad \text{where} \qquad Z_k \coloneqq
 \left ( E_{k,t}^{2}-\noisevar_{k,t}^{2}  \right ) \big[
  \bU_{i,\cdot}^\star(\bU_1^{\star\top}\bU_1^\star)^{-1}\bU_{k,\cdot}^{\star\top}
  \big]^{2}.
\end{align*}
For each $k \in [N_1]$, we can check that $Z_{k}$ is a mean-zero,
sub-exponential random variable (see e.g., \citet[Section
  2.7]{vershynin2016high} for the definition) with $\psi_1$-norm upper
bounded by
\begin{align*}
 \left  \Vert  Z_{k} \right  \Vert  _{\psi_1} \leq \tilde{c} \big[
  \bU_{i,\cdot}^\star(\bU_1^{\star\top}\bU_1^\star)^{-1}\bU_{k,\cdot}^{\star\top}
  \big]^{2}\noisevar_{k,t}^{2}\eqqcolon L_{k},
\end{align*}
where $\tilde{c}>0$ is some universal constant. 

Invoking the Bernstein inequality \citep[Theorem
  2.8.1]{vershynin2016high} guarantees that
\begin{align*}
   \left |\xi_1 \right | & \leq \widetilde{C}
   \sqrt{\sum_{k=1}^{N_1}L_{k}^{2}\log(N+T)} + \widetilde{C} \max_{1
     \leq k \leq N_1} L_{k} \log(N+T)\\ & \overset{\text{(i)}}{\leq}
   \widetilde{C} \noisevar_{\max}\max_{1 \leq k \leq N_1}
   \big|\bU_{i,\cdot}^\star(\bU_1^{\star\top}\bU_1^\star)^{-1}\bU_{k,\cdot}^{\star\top}\big|
   \sqrt{\alpha_{i,t}\spicy}+ \widetilde{C}
   \noisevar_{\max}^{2}\max_{1 \leq k \leq
     N_1}\big[\bU_{i,\cdot}^\star(\bU_1^{\star\top}\bU_1^\star)^{-1}\bU_{k,\cdot}^{\star\top}\big]^{2}\spicy
   \\
& \overset{\text{(ii)}}{\leq} \frac{\widetilde{C}}{\clow}
  \noisevar_{\max} \sqrt{\alpha_{i,t} \spicy}\sqrt{\frac{\mu
      r}{N_1}} \left  \Vert  \bU_{i,\cdot}^\star \right  \Vert  _2 +
  \frac{\widetilde{C}}{\clow^2} \noisevar_{\max}^{2}\frac{\mu
    r}{N_1} \left  \Vert  \bU_{i,\cdot}^\star \right  \Vert  _2^{2} \spicy\\ &
  \overset{\text{(iii)}}{\leq} \frac{\widetilde{C} \ratio
    \cupper^{1/2} }{\clow} \sqrt{\frac{\mu r \spicy}{N}} \alpha_{i,t}
  + \frac{\widetilde{C} \cupper \ratio^2}{\clow^2} \frac{\mu
    r}{N}\spicy \alpha_{i,t} \overset{\text{(iv)}}{\leq}
  \frac{\delta}{16\sqrt{\spicy}} \alpha_{i,t}
\end{align*}
with probability at least $1-\smallprob$, for some universal constant
$\widetilde{C} > 0$. Here step (i) and (iii) both follow from
equation~\eqref{eq:var-lb}; step (ii) utilizes the
relation~\eqref{eq:proof-U}; step (iv) holds under the conditions of
\Cref{thm:CI}. To bound $\xi_2$, we need the following result, whose
proof is deferred to the end of this section.

\begin{claim}
  \label{claim:1}
Under the conditions of Theorem~\ref{thm:CI}, with probability at
least $1-\smallprob$, we have
\begin{align*}
  \left |\delta_{i,t} \right |\leq C_\delta \noisevar_{i,t}^{2}
  \sqrt{\frac{\mu r}{\min\{N_{1},T_{1}\}}} \spicy.
\end{align*}
for some universal constant $C_\delta > 0$.
\end{claim}
Then we can bound $\xi_2$ as follows:
\begin{align*}
\xi_2 & =\sum_{k=1}^{N_1}\delta_{k,t} \left [
  \bU_{i,\cdot}^\star(\bU_1^{\star\top}\bU_1^\star)^{-1}\bU_{k,\cdot}^{\star\top}
  \right ]^{2} \overset{\text{(i)}}{\leq} C_\delta \sqrt{\frac{\mu
    r}{\min\{N_{1},T_{1}\}}} \spicy \sum_{k=1}^{N_1}\noisevar_{k,t}^2
\left [
  \bU_{i,\cdot}^\star(\bU_1^{\star\top}\bU_1^\star)^{-1}\bU_{k,\cdot}^{\star\top}
  \right ]^{2} \\
& \overset{\text{(ii)}}{\leq} C_\delta \sqrt{\frac{\mu
    r\spicy^{2}}{\min\{N_1,T_1\}}} \alpha_{i,t}
\overset{\text{(iii)}}{\leq} \frac{\delta}{16\sqrt{\spicy}}
\alpha_{i,t}
\end{align*}
where step (i) utilizes Claim~\ref{claim:1}; step (ii) follows from
\eqref{eq:var-lb}; while step (iii) holds provided that
$\min\{N_1,T_1\}\gg \delta^{-2}\mu r \spicy^2$.  Taking the bounds on
$\xi_1$ and $\xi_2$ collectively yields
\begin{equation}
\label{eq:proof-var-est-1}  
\left |\overline{u}_{i,t}-u_{i,t}^\star \right | \leq \xi_1+\xi_2\leq
\frac{\delta}{8\sqrt{\spicy}} \alpha_{i,t}.
\end{equation}
Finally, we are left with bounding
$\vert\widehat{u}_{i,t}-\overline{u}_{i,t}\vert$. In view of
Claim~\ref{claim:1} and the sub-Gaussianity of the noise $E_{k,t}$,
with probability at least $1-O((N+T)^{-10}$, we have
\begin{align*}
\widehat{E}_{k,t}^{2}=E_{k,t}^{2}+\delta_{k,t} \leq \widetilde{C} \noisevar_{k,t}^{2}\spicy+C_\delta \noisevar_{k,t}^{2} \sqrt{\frac{\mu r}{\min\{N_{1},T_{1}\}}} \spicy \leq 2\widetilde{C} \noisevar_{k,t}^{2}\spicy,
\end{align*}
provided that $\min\{N_1,T_1\}\gg \mu r \spicy^2$.
Hence we have
\begin{align}
	 \left |\widehat{u}_{i,t}-\overline{u}_{i,t} \right | & =\sum_{k=1}^{N_1}\widehat{E}_{k,t}^{2} \left |\big[\widehat{\bU}_{i,\cdot}(\widehat{\bU}_1^{\top}\widehat{\bU}_1)^{-1}\widehat{\bU}_{k,\cdot}^{\top}\big]^{2}- \left [ \bU_{i,\cdot}^\star(\bU_1^{\star\top}\bU_1^\star)^{-1}\bU_{k,\cdot}^{\star\top}  \right ]^{2} \right |\nonumber \\
	& \leq \widetilde{C} \spicy \sum_{k=1}^{N_1}\noisevar_{k,t}^{2} \Big|\big[\widehat{\bU}_{i,\cdot}(\widehat{\bU}_1^{\top}\widehat{\bU}_1)^{-1}\widehat{\bU}_{k,\cdot}^{\top}\big]^{2}-\big[ \bU_{i,\cdot}^\star(\bU_1^{\star\top}\bU_1^\star)^{-1}\bU_{k,\cdot}^{\star\top} \big]^{2}\Big|\nonumber \\
	& \leq  \frac{\delta}{8\sqrt{\spicy}}\alpha_{i,t},\label{eq:proof-var-est-2}
\end{align}
where the penultimate relation follows from the following Claim~\ref{claim:2}.
\begin{claim} \label{claim:2}
Under the conditions of \Cref{thm:CI}, for each $N_1 < i\leq N$ and $1\leq k \leq N_1$, with probability at least $1-O((N+T)^{-10}$ we have
\begin{align*}
\sum_{k=1}^{N_1}\noisevar_{\max}^{2} \Big|\big[\widehat{\bU}_{i,\cdot}(\widehat{\bU}_1^{\top}\widehat{\bU}_1)^{-1}\widehat{\bU}_{k,\cdot}^{\top}\big]^{2}-\big[ \bU_{i,\cdot}^\star(\bU_1^{\star\top}\bU_1^\star)^{-1}\bU_{k,\cdot}^{\star\top} \big]^{2}\Big| \leq \frac{\delta}{8\widetilde{C}\spicy^{3/2}} \alpha_{i,t}.
\end{align*}
\end{claim}
The proof of Claim~\ref{claim:2} is deferred to the end of this
section.  Taken together, equations~\eqref{eq:proof-var-est-1}
and~\eqref{eq:proof-var-est-2} imply that
\begin{align*}
 \left |\widehat{u}_{i,t}-u_{i,t}^\star \right | \leq \left
 |\overline{u}_{i,t}-u_{i,t}^\star \right |+ \left
 |\widehat{u}_{i,t}-\overline{u}_{i,t} \right
 |\leq\frac{\delta}{4\sqrt{\spicy}}\alpha_{i,t}.
\end{align*}
Similarly, we can establish that \mbox{$\vert\widehat{v}_{i,t} -
  v_{i,t}^\star\vert \leq \frac{\delta}{4\sqrt{\spicy}}\beta_{i,t}$.}
Therefore, we arrive at the desired bound
\begin{align*}
 \left |\estvar_{i,t} - \estvar_{i,t}^\star  \right | & \leq
 \left |\widehat{u}_{i,t} - u_{i,t}^\star  \right | +
\vert\widehat{v}_{i,t} - v_{i,t}^\star\vert \leq \frac{\delta}{4
  \sqrt{\spicy}}\estvar_{i,t}^\star,
\end{align*}
where we use the fact that $\estvar_{i,t}^\star=\alpha_{i,t}+\beta_{i,t}$.


\paragraph{Proof of Claim~\ref{claim:1}.} 
Recall the definition~\eqref{eq:defn-delta} of $\delta_{i,t}$ and the
decomposition
$\widehat{M}_{i,t}=M_{i,t}^{\star}+Z_{i,t}+\Delta_{i,t}$.  With this
notation, we have
\begin{align*}
\delta_{i,t} & \coloneqq
\big(M_{i,t}^{\star}-\widehat{M}_{i,t}\big)^{2} + 2 E_{i,t}
\big(M_{i,t}^{\star}-\widehat{M}_{i,t} \big) = (Z_{i,t} +
\Delta_{i,t})^{2} + 2 E_{i,t} \big(Z_{i,t}+\Delta_{i,t} \big).
\end{align*}
Then we can bound $\delta_{i,t}$ as follows:
\begin{align*}
\vert\delta_{i,t}\vert & \overset{\text{(i)}}{\leq}
2Z_{i,t}^{2}+2\Delta_{i,t}^{2}+2\vert E_{i,t}\vert\vert
Z_{i,t}\vert+2\vert E_{i,t}\vert\vert\Delta_{i,t}\vert\\ &
\overset{\text{(ii)}}{\leq}
2\widetilde{C}^{2}\gamma_{i,t}^{\star}\spicy+2C_{\Delta}^{2}\delta^{2}\gamma_{i,t}^{\star}+2\widetilde{C}\noisevar_{i,t}\sqrt{\spicy}\cdot\widetilde{C}\sqrt{\gamma_{i,t}^{\star}\spicy}+2\noisevar_{i,t}\sqrt{\spicy}\cdot
C_{\Delta}\delta\sqrt{\gamma_{i,t}^{\star}}\\ &
\overset{\text{(iii)}}{\leq} C_\delta \noisevar_{i,t}^{2}
\sqrt{\frac{\mu r}{\min\{N_{1},T_{1}\}}} \spicy
\end{align*}
as claimed. Here step (i) utilizes the AM-GM inequality; step (ii)
follows from equations~\eqref{eq:proof-CI-3}
and~\eqref{eq:proof-normality-5}, as well as the bound
$|E_{i,t}|\leq\widetilde{C}\noisevar_{i,t}\sqrt{\spicy}$, which holds
with probability at least $1-O((N+T)^{-10})$ for sub-Gaussian noise
$E_{i,t}$; step (iii) follows from the fact that
\begin{align*}
\gamma_{i,t}^\star = \alpha_{i,t}+\beta_{i,t}\leq
\clow^{-1}\frac{N}{N_1}\noisevar_{\max}^{2} \left \Vert
\bU_{i,\cdot}^\star \right \Vert _2^{2} +
\clow^{-1}\frac{T}{T_1}\noisevar_{\max}^{2} \left \Vert
\bV_{t,\cdot}^\star \right \Vert _2^{2} \leq 2 \clow^{-1}
\noisevar_{\max}^2 \frac{\mu r}{\min\{N_1,T_1\}},
\end{align*}and holds provided that $C_\delta \geq  3 \widetilde{C}^2 + 3 C_\Delta + \kappa_{\sigma}$ and under the conditions of~\Cref{thm:CI}.


\paragraph{Proof of Claim~\ref{claim:2}.}

We first record a few results from the paper~\cite{yan2024entrywise}:
there exists a rotation matrix $\bH \in \RR^{r\times r}$ and some
universal constant $\widetilde{c}>0$, such that with probability at
least $1-O((N+T)^{-10}$,
\begin{subequations}
\begin{equation}
\big  \Vert (\widehat{\bU}_{1}\bH)^{\top} \widehat{\bU}_{1} \bH
\big)^{-1} - (\bU_{1}^{\star\top}\bU_{1}^{\star})^{-1} \big  \Vert  \leq
\widetilde{c} \frac{N^{2}}{N_{1}^{2}} \Big(\frac{\noisevar_{\max}
  \sqrt{r + \spicy}}{\sing_{r}^{\star}
  \sqrt{T_{1}/T}}\sqrt{\frac{N_{1}}{N}} + \frac{\noisevar_{\max}^{2}(N
  + T_{1})}{\sing_{r}^{\star2}T_{1}/T} \frac{N_{1}}{N}
\Big).\label{eq:proof-claim-2-1}
\end{equation}
As an immediate consequence, under the conditions of \Cref{thm:CI}, we
have
\begin{equation}
\label{eq:proof-claim-2-2}  
 \Vert (\widehat{\bU}_{1}^{\top}\widehat{\bU}_{1})^{-1} \Vert
 \stackrel{(a)}{\leq} 2 \clow^{-1} \frac{N}{N_{1}}, \quad \mbox{and}
 \quad \big \Vert \widehat{\bU}_{k,\cdot}\big \Vert _{2}
 \stackrel{(b)}{\leq} 2 \Vert \bU_{k,\cdot}^{\star} \Vert _{2} +
 \widetilde{c} \frac{\noisevar_{\max}\sqrt{r +
     \spicy}}{\sing_{r}^{\star} \sqrt{T_{1}/T}},
\end{equation}
Moreover, under the conditions of~\Cref{thm:CI}, we have
\begin{equation}
\label{eq:proof-claim-2-4}  
\big \Vert \widehat{\bU}_{i,\cdot}\big \Vert _{2} \leq 3 \Vert
\bU_{i,\cdot}^{\star} \Vert _{2}.
\end{equation}
In addition, for each $k \in [N]$ we have
\begin{equation}
  \big  \Vert  \widehat{\bU}_{k,\cdot} \bH -\bU_{k,\cdot}^{\star}\big
   \Vert _{2} \leq \widetilde{c} \frac{\noisevar_{\max} \sqrt{r +
      \spicy}}{\sing_{r}^{\star} \sqrt{T_{1}/T}} +
  \widetilde{c}\frac{\noisevar_{\max}^{2}(N+T_{1})}{\sing_{r}^{\star2}T_{1}/T} \Vert \bU_{k,\cdot}^{\star} \Vert _{2}.\label{eq:proof-claim-2-5}
\end{equation}
\end{subequations}
Here equations~\eqref{eq:proof-claim-2-1},
\eqref{eq:proof-claim-2-2}(b) and \eqref{eq:proof-claim-2-5} come from
equations (B.13), (B.17), (B.4) in the paper~\cite{yan2024entrywise}.

\newcommand{\Hack}{\ensuremath{\mathbf{\eta}}}

To establish Claim~\ref{claim:2}, we introduce the shorthand $\Hack
\defn \widehat{\bU}_{i,\cdot}
(\widehat{\bU}_{1}^{\top}\widehat{\bU}_{1})^{-1}\widehat{\bU}_{k,\cdot}^{\top}-\bU_{i,\cdot}^{\star}(\bU_{1}^{\star\top}\bU_{1}^{\star})^{-1}\bU_{k,\cdot}^{\star\top}$, and
the decomposition $\Hack = \sum_{j=1}^3 \theta_j$, where
\begin{align*}
\theta_1 & \defn \widehat{\bU}_{i,\cdot} \bH
\big[\big((\widehat{\bU}_{1}\bH)^{\top}\widehat{\bU}_{1}\bH\big)^{-1}-(\bU_{1}^{\star\top}\bU_{1}^{\star})^{-1}\big]
(\widehat{\bU}_{k,\cdot}\bH)^{\top} \\ 
\theta_2 & \defn \big(\widehat{\bU}_{i,\cdot} \bH -
\bU_{i,\cdot}^{\star}\big)(\bU_{1}^{\star\top}\bU_{1}^{\star})^{-1}(\widehat{\bU}_{k,\cdot}\bH)^{\top} \quad \mbox{and} \\
\theta_3 & \defn
\bU_{i,\cdot}^{\star}(\bU_{1}^{\star\top}\bU_{1}^{\star})^{-1}(\widehat{\bU}_{k,\cdot}\bH-\bU_{k,\cdot}^{\star})^{\top}.
\end{align*}
We bound each of the quantities $\{|\theta_j|\}_{j=1}^3$ in turn.
We have
\begin{align*}
  \vert \theta_{1}\vert & \leq \big \Vert \widehat{\bU}_{i,\cdot}
  \Vert _{2}\big \Vert
  (\widehat{\bU}_{1}\bH)^{\top}\widehat{\bU}_{1}\bH\big)^{-1}-(\bU_{1}^{\star\top}\bU_{1}^{\star})^{-1}\big
  \Vert \big \Vert \widehat{\bU}_{k,\cdot} \Vert _{2}\\ &
  \overset{\text{(i)}}{\leq}\widetilde{c}\frac{N^{2}}{N_{1}^{2}}\Big(\frac{\noisevar_{\max}\sqrt{r+\spicy}}{\sing_{r}^{\star}\sqrt{T_{1}/T}}\sqrt{\frac{N_{1}}{N}}+\frac{\noisevar_{\max}^{2}(N+T_{1})}{\sing_{r}^{\star2}T_{1}/T}\frac{N_{1}}{N}\Big)
  3 \Vert \bU_{i,\cdot}^{\star} \Vert _{2} \Big(2 \Vert
  \bU_{k,\cdot}^{\star} \Vert
  _{2}+\widetilde{c}\frac{\noisevar_{\max}\sqrt{r+\spicy}}{\sing_{r}^{\star}\sqrt{T_{1}/T}}\Big),
\end{align*}
where step (i) follows from (\ref{eq:proof-claim-2-1}). The term
$|\theta_2|$ can be bounded by
\begin{align*}
	\vert\theta_{2}\vert & \leq\clow^{-1}\frac{N}{N_{1}}\big \Vert
        \widehat{\bU}_{i,\cdot}\bH-\bU_{i,\cdot}^{\star}\big \Vert
        _{2}\big \Vert \widehat{\bU}_{k,\cdot} \Vert _{2}\\ &
        \overset{\text{(ii)}}{\leq}\frac{\widetilde{c}}{\clow}\frac{N}{N_{1}}\Big(\frac{\noisevar_{\max}\sqrt{r+\spicy}}{\sing_{r}^{\star}\sqrt{T_{1}/T}}+\frac{\noisevar_{\max}^{2}(N+T_{1})}{\sing_{r}^{\star2}T_{1}/T}
        \Vert \bU_{i,\cdot}^{\star} \Vert _{2}\Big)\Big(2 \Vert
        \bU_{k,\cdot}^{\star} \Vert
        _{2}+\widetilde{c}\frac{\noisevar_{\max}\sqrt{r+\spicy}}{\sing_{r}^{\star}\sqrt{T_{1}/T}}\Big),
\end{align*}
where step (ii) follows from the bounds~\eqref{eq:proof-claim-2-2}(b); 
and~\eqref{eq:proof-claim-2-5}. Finally, the term $|\theta_3|$ can be
bounded by
\begin{align*}
  |\theta_{3}\vert & \overset{\text{(iii)}}{\leq}
  \clow^{-1}\frac{N}{N_{1}} \Vert \bU_{i,\cdot}^{\star} \Vert
  _{2}\Big(\widetilde{c}\frac{\noisevar_{\max}\sqrt{r+\spicy}}{\sing_{r}^{\star}\sqrt{T_{1}/T}}+\widetilde{c}\frac{\noisevar_{\max}^{2}(N+T_{1})}{\sing_{r}^{\star2}T_{1}/T}
  \Vert \bU_{k,\cdot}^{\star} \Vert _{2}\Big)\\ &
  =\frac{\widetilde{c}}{\clow}\frac{N}{N_{1}} \Vert
  \bU_{i,\cdot}^{\star} \Vert
  _{2}\frac{\noisevar_{\max}\sqrt{r+\spicy}}{\sing_{r}^{\star}\sqrt{T_{1}/T}}+\frac{\widetilde{c}}{\clow}\frac{N}{N_{1}}\frac{\noisevar_{\max}^{2}(N+T_{1})}{\sing_{r}^{\star2}T_{1}/T}
  \Vert \bU_{i,\cdot}^{\star} \Vert _{2} \Vert \bU_{k,\cdot}^{\star}
  \Vert _{2}.
\end{align*}
where step (iii) follows from equation~\eqref{eq:proof-claim-2-5}).
Since $\Hack \leq \sum_{j=1}^3 |\theta_j|$, collecting the the above
three bounds yields
\begin{subequations}
\begin{align}
\Hack & \leq \frac{N}{N_{1}} \Vert \bU_{i,\cdot}^{\star} \Vert _{2}
\Vert \bU_{k,\cdot}^{\star} \Vert _{2}\Big(4\widetilde{c}
\frac{\noisevar_{\max}\sqrt{r+\spicy}}{\sing_{r}^{\star}}\sqrt{\frac{NT}{N_{1}T_{1}}}+7
\frac{\widetilde{c}}{\clow}\frac{\noisevar_{\max}^{2}(N+T_{1})}{\sing_{r}^{\star2}T_{1}/T}\Big)
\nonumber \\
& \qquad \qquad + 2\frac{\widetilde{c}}{\clow}\frac{N}{N_{1}} \big(
\Vert \bU_{i,\cdot}^{\star} \Vert _{2} + \Vert \bU_{k,\cdot}^{\star}
\Vert _{2}
\big)\frac{\noisevar_{\max}\sqrt{r+\spicy}}{\sing_{r}^{\star}\sqrt{T_{1}/T}}
+
\frac{\widetilde{c}^2}{\clow}\frac{N}{N_{1}}\Big(\frac{\noisevar_{\max}\sqrt{r+\spicy}}{\sing_{r}^{\star}\sqrt{T_{1}/T}}\Big)^2,\label{eq:proof-claim-2-6}
\end{align}
provided that $\SNR \leq c_0/\sqrt{r+\spicy}$ for some sufficiently
small constant $c_0>0$.  Continuing the argument, the quantity $\Hack'
\defn
\big\vert\widehat{\bU}_{i,\cdot}(\widehat{\bU}_{1}^{\top}\widehat{\bU}_{1})^{-1}\widehat{\bU}_{k,\cdot}^{\top}+\bU_{i,\cdot}^{\star}(\bU_{1}^{\star\top}\bU_{1}^{\star})^{-1}\bU_{k,\cdot}^{\star\top}\big\vert$
is bounded as
\begin{align}
  \Hack' & \overset{\text{(a)}}{\leq}2\clow^{-1}\frac{N}{N_{1}}\big \Vert
\widehat{\bU}_{i,\cdot}\big \Vert _{2}\big \Vert
\widehat{\bU}_{k,\cdot}\big \Vert _{2}+\clow^{-1}\frac{N}{N_{1}} \Vert
\bU_{i,\cdot}^{\star} \Vert _{2} \Vert \bU_{k,\cdot}^{\star} \Vert
_{2}\nonumber \\
& \overset{\text{(b)}}{\leq}6\clow^{-1}\frac{N}{N_{1}} \Vert
\bU_{i,\cdot}^{\star} \Vert _{2}\Big(2 \Vert \bU_{k,\cdot}^{\star}
\Vert
_{2}+\widetilde{c}\frac{\noisevar_{\max}\sqrt{r+\spicy}}{\sing_{r}^{\star}\sqrt{T_{1}/T}}\Big)+\clow^{-1}\frac{N}{N_{1}}
\Vert \bU_{i,\cdot}^{\star} \Vert _{2} \Vert \bU_{k,\cdot}^{\star}
\Vert _{2}\nonumber \\
\label{eq:proof-claim-2-7}
& \leq 13 \clow^{-1}\frac{N}{N_{1}} \Vert \bU_{i,\cdot}^{\star} \Vert
_{2} \Vert \bU_{k,\cdot}^{\star} \Vert
_{2}+6\frac{\widetilde{c}}{\clow}\frac{N}{N_{1}}\frac{\noisevar_{\max}\sqrt{r+\spicy}}{\sing_{r}^{\star}\sqrt{T_{1}/T}}
\Vert \bU_{i,\cdot}^{\star} \Vert _{2}.
\end{align}
\end{subequations}
Here step (a) follows from the bound~\eqref{eq:proof-claim-2-2}(a),
while step (b) follows from the bounds~\eqref{eq:proof-claim-2-2}(b)
and~\eqref{eq:proof-claim-2-4}.

Combining the bounds~\eqref{eq:proof-claim-2-6}
and~\eqref{eq:proof-claim-2-7} yields
\begin{align*}
&
  \big\vert\big(\widehat{\bU}_{i,\cdot}(\widehat{\bU}_{1}^{\top}\widehat{\bU}_{1})^{-1}\widehat{\bU}_{k,\cdot}^{\top}\big)^{2}-\big(\bU_{i,\cdot}^{\star}(\bU_{1}^{\star\top}\bU_{1}^{\star})^{-1}\bU_{k,\cdot}^{\star\top}\big)^{2}\big\vert
  \\ &
  \qquad\leq13\clow^{-1}\frac{N^{2}}{N_{1}^{2}} \Vert \bU_{i,\cdot}^{\star} \Vert _{2}^{2} \Vert \bU_{k,\cdot}^{\star} \Vert _{2}^{2}\bigg[4\widetilde{c}\frac{\noisevar_{\max}\sqrt{r+\spicy}}{\sing_{r}^{\star}}\sqrt{\frac{NT}{N_{1}T_{1}}}+7\frac{\widetilde{c}}{\clow}\frac{\noisevar_{\max}^{2}(N+T_{1})}{\sing_{r}^{\star2}T_{1}/T}\bigg]\\ &
  \qquad\quad+27\frac{\widetilde{c}}{\clow^{2}}\frac{N^{2}}{N_{1}^{2}} \Vert \bU_{i,\cdot}^{\star} \Vert _{2}^{2} \Vert \bU_{k,\cdot}^{\star} \Vert _{2}\frac{\noisevar_{\max}\sqrt{r+\spicy}}{\sing_{r}^{\star}\sqrt{T_{1}/T}}+26\frac{\widetilde{c}}{\clow^{2}}\frac{N^{2}}{N_{1}^{2}} \Vert \bU_{i,\cdot}^{\star} \Vert _{2} \Vert \bU_{k,\cdot}^{\star} \Vert _{2}^{2}\frac{\noisevar_{\max}\sqrt{r+\spicy}}{\sing_{r}^{\star}\sqrt{T_{1}/T}}\\ &
  \qquad\quad+37\frac{\widetilde{c}^{2}}{\clow^{2}}\frac{N^{2}}{N_{1}^{2}}\Big(\frac{\noisevar_{\max}\sqrt{r+\spicy}}{\sing_{r}^{\star}\sqrt{T_{1}/T}}\Big)^{2} \Vert \bU_{i,\cdot}^{\star} \Vert _{2} \Vert \bU_{k,\cdot}^{\star} \Vert _{2}+6\frac{\widetilde{c}^{2}}{\clow^{2}}\frac{N^{2}}{N_{1}^{2}} \Vert \bU_{i,\cdot}^{\star} \Vert _{2}\Big(\frac{\noisevar_{\max}\sqrt{r+\spicy}}{\sing_{r}^{\star}\sqrt{T_{1}/T}}\Big)^{3}.
\end{align*}
Summing up the above inequality from $k=1$ to $N_1$ yields
\begin{align*}
	& \sigma_{\max}^2
  \sum_{k=1}^{N_{1}}\big\vert\big(\widehat{\bU}_{i,\cdot}(\widehat{\bU}_{1}^{\top}\widehat{\bU}_{1})^{-1}\widehat{\bU}_{k,\cdot}^{\top}\big)^{2}-\big(\bU_{i,\cdot}^{\star}(\bU_{1}^{\star\top}\bU_{1}^{\star})^{-1}\bU_{k,\cdot}^{\star\top}\big)^{2}\big\vert\\ &
  \quad\overset{\text{(i)}}{\leq}13\frac{\cupper}{\clow}
  \sigma_{\max}^2
  \frac{N^{2}}{N_{1}^{2}} \Vert \bU_{i,\cdot}^{\star} \Vert _{2}^{2}\frac{N_{1}}{N}r\bigg[4\widetilde{c}\frac{\noisevar_{\max}\sqrt{r+\spicy}}{\sing_{r}^{\star}}\sqrt{\frac{NT}{N_{1}T_{1}}}+7\frac{\widetilde{c}}{\clow}\frac{\noisevar_{\max}^{2}(N+T_{1})}{\sing_{r}^{\star2}T_{1}/T}\bigg]\\ &
  \quad\quad+27\frac{\widetilde{c}}{\clow^{2}}\frac{N^{2}}{N_{1}} \Vert \bU_{i,\cdot}^{\star} \Vert _{2}^{2}\sqrt{\frac{\mu
      r}{N}}\frac{\noisevar_{\max}^3\sqrt{r+\spicy}}{\sing_{r}^{\star}\sqrt{T_{1}/T}}+26\frac{\widetilde{c}\cupper}{\clow^{2}}\frac{N^{2}}{N_{1}^{2}} \Vert \bU_{i,\cdot}^{\star} \Vert _{2}r\frac{\noisevar_{\max}^3\sqrt{r+\spicy}}{\sing_{r}^{\star}\sqrt{T_{1}/T}}\\ &
  \quad\quad+40\frac{\widetilde{c}^{2}}{\clow^{2}}\frac{N^{2}}{N_{1}}\Big(\frac{\noisevar_{\max}^2\sqrt{r+\spicy}}{\sing_{r}^{\star}\sqrt{T_{1}/T}}\Big)^{2} \Vert \bU_{i,\cdot}^{\star} \Vert _{2}\sqrt{\frac{\mu
      r}{N}} \\
& \overset{\text{(ii)}}{\leq}
  \frac{\delta}{8\widetilde{C}\spicy^{3/2}} \alpha_{i,t}.
\end{align*}
Here step (i) follows from the bounds
\begin{align*}
\sum_{k=1}^{N_{1}} \Vert \bU_{k,\cdot}^{\star} \Vert _{2}^{2} & =
\Vert \bU_{1}^{\star} \Vert _{\mathrm{F}}^{2}\leq r \Vert
\bU_{1}^{\star} \Vert ^{2}\leq\cupper\frac{N_{1}}{N} r, \quad
\mbox{and} \quad
\sum_{k=1}^{N_{1}} \Vert \bU_{k,\cdot}^{\star} \Vert _{2} & \leq N_{1}
\Vert \bU^{\star} \Vert _{2, \infty}\leq N_{1}\sqrt{\frac{\mu r}{N}},
\end{align*}
while step (ii) holds under the conditions of~\Cref{thm:CI}.

\section{Inferential procedure for bilinear forms}
\label{AppBilinear}

In this section, we discuss how to estimate and construct confidence
intervals for bilinear forms of the hidden block $\bM_d^\star$ under
the four-block model.  More precisely, given any two fixed vectors
$\lvec\in\mathbb{R}^{N_{2}}$ and $\rvec \in
\mathbb{R}^{T_{2}}$, we describe how to compute a confidence interval
for the quantity $\lvec^{\top}\bM_{d}^{\star}\rvec$.

From our discussion following~\Cref{thm:CI}, we have the approximation
$\Mhat_d - \bM_d^\star \approx \bZ$, where the matrix $\bZ$ was
defined in equation~\eqref{eq:Z-defn}.  Our previous analysis
established this approximation in an entrywise; similar arguments can
be used to show that
$\lvec^{\top}(\Mhat_{d}^{\star}-\bM_d^\star)\rvec\approx
\lvec^{\top}\bZ\rvec$.  In this way, we find that
\begin{equation}
\label{eq:bilinear-approx}
\lvec^{\top}\big(\Mhat_{d}-\bM_{d}^{\star} \big)\rvec
\approx\big\langle \bE_{b},\bU_{1}^{\star} (\bU_{1}^{\star\top}
\bU_{1}^{\star} )^{-1}\bU_{2}^{\star\top} \lvec
\rvec^{\top}\big\rangle +\big\langle \bE_{c},\lvec
\rvec^{\top}\bV_{2}^{\star} (\bV_{1}^{\star\top} \bV_{1}^{\star}
)^{-1}\bV_{1}^{\star\top} \big\rangle.
\end{equation}
The right hand side is a linear form of independent noise components,
which follows approximately a mean-zero Gaussian distribution with
variance
\begin{subequations}
\begin{align}
\label{eq:defn-variance-bilinear}
\underbrace{\sum_{i=1}^{N_{1}}\sum_{t=1}^{T_{2}}\noisevar_{i,T_{1}+t}^{2}\big[\bU_{1}^{\star}
    (\bU_{1}^{\star\top} \bU_{1}^{\star} )^{-1}\bU_{2}^{\star\top}
    \lvec \rvec^{\top}\big]_{i,t}^{2}
  +\sum_{i=1}^{N_{2}}\sum_{t=1}^{T_{1}}\noisevar_{N_{1}+i,t}^{2}\big[\lvec
    \rvec^{\top}\bV_{2}^{\star} (\bV_{1}^{\star\top} \bV_{1}^{\star}
    )^{-1}\bV_{1}^{\star\top} \big]_{i,t}^{2}}_{\eqqcolon
  \estvar^{\star} (\lvec,\rvec)}.
\end{align}
A natural data-driven estimate
$\widehat{\estvar}(\lvec,\rvec)$ is given by
\begin{align}
\label{eq:gamma-hat-bilinear}
\sum_{i=1}^{N_{1}}\sum_{t=1}^{T_{2}}
\widehat{E}_{i,T_{1}+t}^{2}\big[\widehat{\bU}_{1}
  (\widehat{\bU}_{1}^{\top}
  \widehat{\bU}_{1})^{-1}\widehat{\bU}_{2}^{\top} \lvec
  \rvec^{\top}\big]_{i,t}^{2} + \sum_{i=1}^{N_{2}}\sum_{t=1}^{T_{1}}
\widehat{E}_{N_{1}+i,t}^{2}\big[\lvec \rvec^{\top}\widehat{\bV}_{2}
  (\widehat{\bV}_{1}^{\top}
  \widehat{\bV}_{1})^{-1}\widehat{\bV}_{1}^{\top} \big]_{i,t}^{2}.
  \end{align}
\end{subequations}
This estimation is motivated by the same rationale leading to the
construction of $\widehat{\estvar}_{i,t}$ for Algorithm
\FourBlockConf.  This paves the way for constructing
$(1-\alpha)$-confidence interval for $\lvec^{\top} \bM_{d}^{\star}
\rvec$ as follows
\begin{equation*}
\mathsf{CI}^{(1-\alpha)}_{\lvec,\rvec} \coloneqq \Big[\lvec^\top
  \Mhat_d \rvec \pm \Phi^{-1} (1 - \alpha/2)
  \sqrt{\widehat{\estvar}(\lvec,\rvec)}
\end{equation*}
where $\Phi$ is the CDF of the standard normal distribution.

The validity of this confidence interval (i.e., results similar
to~\Cref{thm:CI}) can be established following the similar analysis as
in the current paper under slightly different conditions, e.g., the
quantities $ \Vert \bU_{i,\cdot}^\star \Vert _2$ and $ \Vert
\bV_{t,\cdot}^\star \Vert _2$ in condition~\eqref{eq:signal} should be
changed to $ \Vert \lvec^\top\bU_2^\star \Vert _2 / \Vert \lvec \Vert
_2$ and $ \Vert \rvec^\top\bV_2^\star \Vert _2 / \Vert \lvec \Vert_2$.

\end{document}